\pdfoutput=1
\documentclass[12pt,a4paper]{article}
\usepackage{ifthen}
\usepackage{booktabs}
\usepackage{overpic}
\usepackage{relsize}
\usepackage{rotating}
\usepackage{enumitem}
\usepackage[top=1in, bottom=1.25in, left=1in, right=1in]{geometry}

\newboolean{pdflatex}
\setboolean{pdflatex}{true}

\newboolean{articletitles}
\setboolean{articletitles}{true}

\newboolean{uprightparticles}
\setboolean{uprightparticles}{false}

\newboolean{inbibliography}
\setboolean{inbibliography}{false}

\def\paperauthors{LHCb collaboration}
\def\paperasciititle{Search for dark photons decaying into muons}
\def\papertitle{Search for \atomm decays}
\def\paperkeywords{{High Energy Physics}, {LHCb}}
\def\papercopyright{CERN on behalf of the LHCb collaboration}
\def\paperlicence{CC-BY-4.0}
\def\paperlicenceurl{https://creativecommons.org/licenses/by/4.0/}

\usepackage{microtype}
\usepackage{lineno}
\usepackage{xspace}
\usepackage{caption}

\usepackage{graphicx}
\usepackage{color}
\usepackage{colortbl}
\graphicspath{{./figs/}}

\usepackage{amsmath}
\usepackage{amssymb}
\usepackage{amsfonts}
\usepackage{upgreek}

\newcommand*\patchAmsMathEnvironmentForLineno[1]{%
\expandafter\let\csname old#1\expandafter\endcsname\csname #1\endcsname
\expandafter\let\csname oldend#1\expandafter\endcsname\csname
end#1\endcsname
 \renewenvironment{#1}%
   {\linenomath\csname old#1\endcsname}%
   {\csname oldend#1\endcsname\endlinenomath}%
}
\newcommand*\patchBothAmsMathEnvironmentsForLineno[1]{%
  \patchAmsMathEnvironmentForLineno{#1}%
  \patchAmsMathEnvironmentForLineno{#1*}%
}
\AtBeginDocument{%
\patchBothAmsMathEnvironmentsForLineno{equation}%
\patchBothAmsMathEnvironmentsForLineno{align}%
\patchBothAmsMathEnvironmentsForLineno{flalign}%
\patchBothAmsMathEnvironmentsForLineno{alignat}%
\patchBothAmsMathEnvironmentsForLineno{gather}%
\patchBothAmsMathEnvironmentsForLineno{multline}%
\patchBothAmsMathEnvironmentsForLineno{eqnarray}%
}

\usepackage{hyperxmp}

\usepackage[pdftex,
            pdfauthor={\paperauthors},
            pdftitle={\paperasciititle},
            pdfkeywords={\paperkeywords},
            pdfcopyright={Copyright (C) \papercopyright},
            pdflicenseurl={\paperlicenceurl}]{hyperref}

\usepackage[all]{hypcap} % Internal hyperlinks to floats.

\usepackage{cite}
\usepackage{mciteplus}

\def\lhcb {\mbox{LHCb}\xspace}

\newcommand{\tev}{\ifthenelse{\boolean{inbibliography}}{\ensuremath{~T\kern -0.05em eV}}{\ensuremath{\mathrm{\,Te\kern -0.1em V}}}\xspace}
\newcommand{\gev}{\ensuremath{\mathrm{\,Ge\kern -0.1em V}}\xspace}
\newcommand{\mev}{\ensuremath{\mathrm{\,Me\kern -0.1em V}}\xspace}
\def\invfb   {\ensuremath{\mbox{\,fb}^{-1}}\xspace}
\def \xip {\ensuremath{\chi^2_{\mathrm{IP}}(\mu)}\xspace}
\def \mxip {\ensuremath{{\rm min}[\chi^2_{\mathrm{IP}}(\mu^{\pm})]}\xspace}

\def\aprime{\ensuremath{A^{\prime}}\xspace}
\def\atomm{\ensuremath{\aprime\!\to\!\mu^+\mu^-}\xspace}
\def\gtomm{\ensuremath{\gamma^*\!\to\!\mu^+\mu^-}\xspace}
\def \mmm {\ensuremath{m(\mu^+\mu^-)}\xspace}
\def\ps   {\ensuremath{\mbox{\,ps}}\xspace}
\def\ma{\ensuremath{m(\aprime)}\xspace}
\def\ta{\ensuremath{\tau(\aprime)}\xspace}
\def\pt         {\mbox{$p_{\rm T}$}\xspace}
\def\KS      {{\ensuremath{K^0_{\mathrm{ \scriptscriptstyle S}}}}\xspace}

\newcommand{\chisq}{\ensuremath{\chi^2}\xspace}
\def\sa{\ensuremath{\sigma[\mmm]}\xspace}
\def\jpsi     {{\ensuremath{{J\mskip -3mu/\mskip -2mu\psi\mskip 2mu}}}\xspace}
\mathchardef\Upsilon="7107

\mathchardef\PLambda="7103
\def\ngob {\ensuremath{n_{\rm ob}^{\gamma^*}[\ma]}\xspace}
\def\naob {\ensuremath{n_{\rm ob}^{\aprime}[\ma]}\xspace}
\def\naex {\ensuremath{n_{\rm ex}^{\aprime}[\ma,\varepsilon^2]}\xspace}
\def\naobt {\ensuremath{n_{\rm ob}^{\aprime}[\ma,\tau(\aprime)]}\xspace}
\def\naobe {\ensuremath{n_{\rm ob}^{\aprime}[\ma,\varepsilon^2]}\xspace}
\def\effr {\ensuremath{\epsilon_{{}^{\gamma^*}}^{{}_{\aprime}}[\ma,\ta]}\xspace}
\def\maeps {\ensuremath{[\ma,\varepsilon^2]}\xspace}

\begin{document}

\renewcommand{\thefootnote}{\fnsymbol{footnote}}
\setcounter{footnote}{1}

\begin{titlepage}
\pagenumbering{roman}

% Header ---------------------------------------------------
\vspace*{-1.5cm}
\centerline{\large EUROPEAN ORGANIZATION FOR NUCLEAR RESEARCH (CERN)}
\vspace*{1.5cm}
\noindent
\begin{tabular*}{\linewidth}{lc@{\extracolsep{\fill}}r@{\extracolsep{0pt}}}
\vspace*{-1.5cm}\mbox{\!\!\!\includegraphics[width=.14\textwidth]{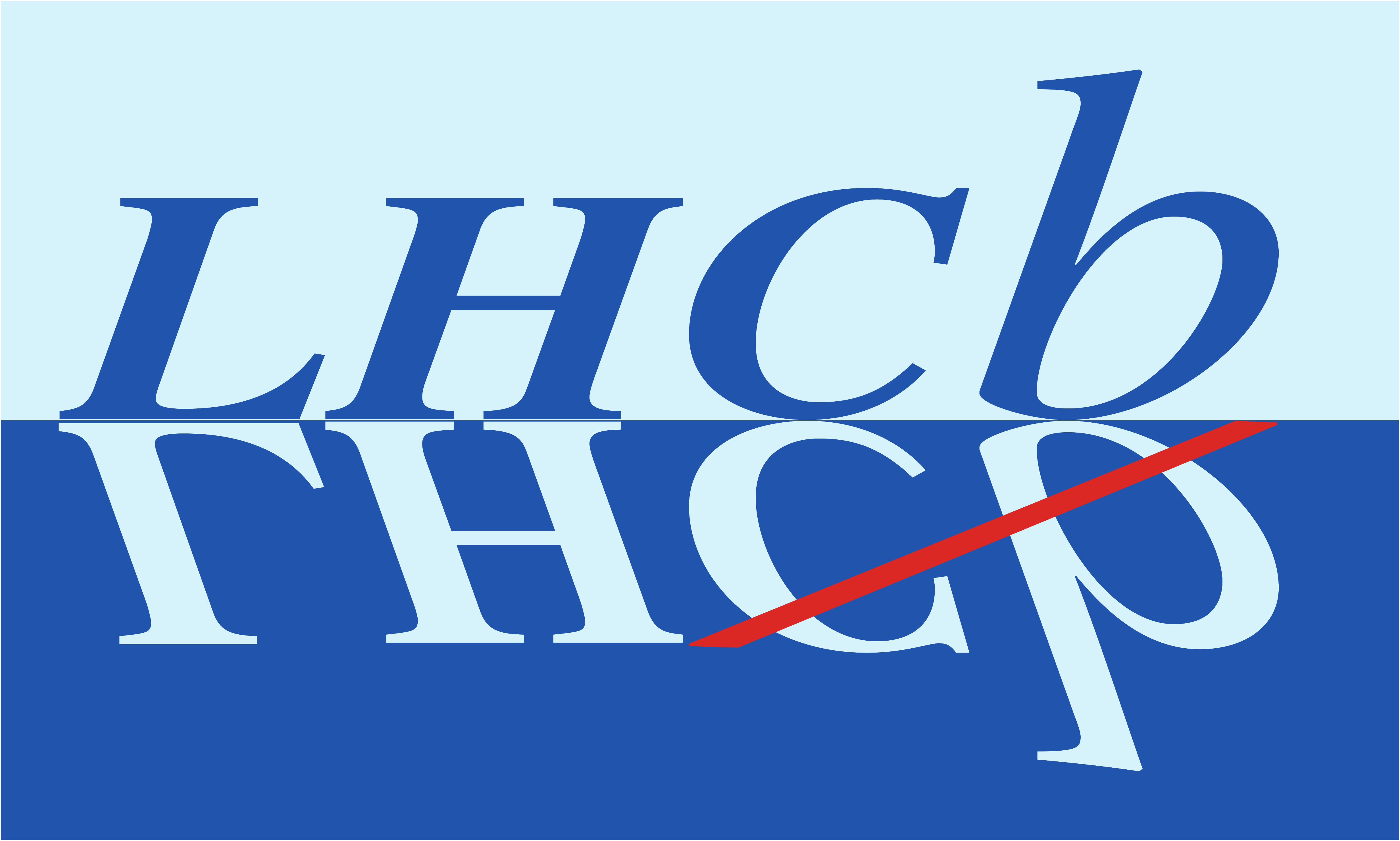}} & & \\
 & & CERN-EP-2019-212 \\  % ID
 & & LHCb-PAPER-2019-031 \\  % ID
 & &  December 22, 2019 \\
 & & \\
\end{tabular*}

\vspace*{4.0cm}

{\normalfont\bfseries\boldmath\huge
\begin{center}
  \papertitle
\end{center}
}

\vspace*{2.0cm}

\begin{center}
\paperauthors\footnote{Authors are listed at the end of this Letter.}
\end{center}

\vspace{\fill}

\begin{abstract}
  \noindent
Searches are performed for both prompt-like and long-lived dark photons, \aprime, produced in proton-proton collisions at a center-of-mass energy of 13\tev.
These searches look for \atomm decays using a data sample corresponding to an integrated luminosity of 5.5\invfb collected with the LHCb detector.
Neither search finds evidence for a signal, and 90\% confidence-level exclusion limits are placed on the $\gamma$--$\aprime$ kinetic-mixing strength.
The prompt-like \aprime search explores the mass region from near the dimuon threshold up to 70\gev, and places the most stringent constraints to date on dark photons with $214 < \ma \lesssim 740\mev$ and $10.6 < \ma \lesssim 30\gev$.
The search for long-lived \atomm decays places world-leading constraints on low-mass dark photons with lifetimes $\mathcal{O}(1)\ps$.
\end{abstract}

\vspace*{2.0cm}

\begin{center}
  Published as Physical Review Letters {\bf 124} (2020) 041801
\end{center}

\vspace{\fill}

{\footnotesize
\centerline{\copyright~\papercopyright, licence \href{\paperlicenceurl}{\paperlicence}.}}
\vspace*{2mm}

\end{titlepage}

\newpage
\setcounter{page}{2}
\pagestyle{empty}
\mbox{~}

\cleardoublepage

\renewcommand{\thefootnote}{\arabic{footnote}}
\setcounter{footnote}{0}

\pagestyle{plain}
\setcounter{page}{1}
\pagenumbering{arabic}
%\linenumbers

Substantial effort has been dedicated recently~\cite{Essig:2013lka,Alexander:2016aln,Battaglieri:2017aum} to searching for the dark photon, \aprime, a hypothetical massive vector boson that could mediate the interactions of dark matter particles\cite{Tulin:2017ara}, similar to how the ordinary photon, $\gamma$, mediates the electromagnetic~(EM) interactions of charged Standard Model (SM) particles.
The dark photon does not couple directly to SM particles; however,
it can obtain a small coupling to the EM current due to kinetic mixing between the SM hypercharge and \aprime field strength tensors~\cite{Fayet:1980rr,Fayet:1980ad,Okun:1982xi,Galison:1983pa,Holdom:1985ag,Pospelov:2007mp,ArkaniHamed:2008qn,Bjorken:2009mm}.
This coupling, which is suppressed relative to that of the photon by a factor labeled $\varepsilon$,
would provide a portal through which dark photons can be produced in the laboratory, and also via which they can decay into visible SM final states.
If the kinetic mixing arises due to processes described by one- or two-loop diagrams containing high-mass particles, possibly even at the Planck scale, then  $10^{-12} \lesssim \varepsilon^2 \lesssim 10^{-4}$ is expected~\cite{Alexander:2016aln}.
Exploring this {\em few-loop} $\varepsilon$ region is one of the most important near-term goals of dark-sector physics.

Dark photons will decay into visible SM particles if
invisible dark-sector decays are kinematically forbidden.
Constraints have been placed on visible \aprime decays by previous beam-dump~\cite{Bergsma:1985is,Konaka:1986cb,Riordan:1987aw,Bjorken:1988as,Bross:1989mp,Davier:1989wz,Athanassopoulos:1997er,Astier:2001ck,Bjorken:2009mm,Essig:2010gu,Williams:2011qb,Blumlein:2011mv,Gninenko:2012eq,Blumlein:2013cua,Banerjee:2018vgk,Andreas:2012mt,Bergsma:1985qz},
fixed-target~\cite{Abrahamyan:2011gv,Merkel:2014avp,Merkel:2011ze,Adrian:2018scb},
collider~\cite{Aubert:2009cp,Curtin:2013fra,Lees:2014xha,Ablikim:2017aab,Anastasi:2015qla,Anastasi:2018azp},
and rare-meson-decay \cite{Bernardi:1985ny,MeijerDrees:1992kd,Archilli:2011zc,Gninenko:2011uv,Babusci:2012cr,Adlarson:2013eza,Agakishiev:2013fwl,Adare:2014mgk,Batley:2015lha,KLOE:2016lwm}
experiments.
These experiments ruled out the few-loop region for dark-photon masses $\ma \lesssim 10\mev$ ($c=1$ throughout this Letter); however, most of the few-loop region at higher masses remains unexplored.
Constraints on invisible \aprime decays can be found in Refs.~\cite{Essig:2013vha,Davoudiasl:2014kua,Banerjee:2016tad,Lees:2017lec,Adler:2001xv,Adler:2004hp,Artamonov:2009sz,Fayet:2006sp,Fayet:2007ua,Fox:2011fx,CortinaGil:2019nuo,Abdallah:2003np,Abdallah:2008aa};
only the visible scenario is considered here.

Many ideas have been proposed to further explore the
\maeps parameter space~\cite{Essig:2010xa,
Freytsis:2009bh,Balewski:2013oza,
Wojtsekhowski:2012zq,
Beranek:2013yqa,
Echenard:2014lma,
Battaglieri:2014hga,
Raggi:2014zpa,
Alekhin:2015byh,
Gardner:2015wea,
Ilten:2015hya,
Curtin:2014cca,
He:2017ord,Kozaczuk:2017per,
Ilten:2016tkc,
Alexander:2017rfd,
He:2017zzr,
Feng:2017uoz,
Nardi:2018cxi,
DOnofrio:2019dcp,
Tsai:2019mtm}.
%For example, Ref.~\cite{Ilten:2016tkc} proposed an inclusive search for \atomm decays with the  LHCb experiment, and showed that such a search will provide sensitivity to large regions of otherwise inaccessible parameter space with the data collected by the end of Run~3 in 2023.
The LHCb collaboration previously performed a search based on the approach proposed in Ref.~\cite{Ilten:2016tkc} using  %Run~2
data corresponding to 1.6\invfb collected in 2016~\cite{LHCb-PAPER-2017-038}.
The constraints placed on prompt-like dark photons, where the dark-photon lifetime  is small compared to the detector resolution, were the most stringent to date for $10.6 < \ma < 70\gev$ and comparable to the best existing limits for $\ma < 0.5\gev$.
The search for long-lived dark photons was the first to achieve sensitivity using a displaced-vertex signature, though only small regions of \maeps  parameter space were excluded.

This Letter presents searches for both prompt-like and long-lived dark photons produced in proton-proton, $pp$, collisions at a center-of-mass energy of 13\tev, looking for \atomm decays using a data sample corresponding to an integrated luminosity of 5.5\invfb  collected with the LHCb detector in 2016--2018.
The strategies employed in these searches are the same as in Ref.~\cite{LHCb-PAPER-2017-038}, though the three-fold increase in integrated luminosity, improved trigger efficiency during 2017--2018 data taking, and improvements in the analysis provide much better sensitivity to dark photons.
The prompt-like \aprime search is performed from near the dimuon threshold up to 70\gev, achieving a factor of 5\,(2) better sensitivity to $\varepsilon^2$ at low\,(high) masses than Ref.~\cite{LHCb-PAPER-2017-038}.
The long-lived \aprime search
is restricted to the mass range $214<\ma<350\mev$, where the data sample potentially has sensitivity, and provides access to much larger regions of \maeps parameter space.

Both the production and decay kinematics of the \atomm and \gtomm processes are identical, since
dark photons produced in $pp$ collisions via $\gamma$--\aprime mixing inherit the production mechanisms of off-shell photons with $m(\gamma^*) = \ma$.
Furthermore, the expected \atomm signal yield  is related to the observed prompt \gtomm yield in a small $\pm\Delta m$ window around \ma, \ngob,
 by~\cite{Ilten:2016tkc}
\begin{equation}
  \label{eq:norm}
\naex = \varepsilon^2 \left[\frac{\ngob}{2\Delta m}\right] \mathcal{F}[\ma]\, \effr,
\end{equation}
where the dark-photon lifetime, \ta,  is a known function of \ma and $\varepsilon^2$,
$\mathcal{F}$ is a known \ma-dependent function,
and \effr is the \ta-dependent ratio of the \atomm and \gtomm detection efficiencies.
For prompt-like dark photons, \atomm decays are experimentally indistinguishable from prompt \gtomm decays, resulting in $\effr = 1$.
This facilitates a fully data-driven search where most experimental systematic effects cancel, since the observed \atomm yields, \naob, can be normalized to \naex to obtain constraints on $\varepsilon^2$ without any knowledge of the detector efficiency or luminosity.
When \ta is larger than the detector decay-time resolution, \atomm decays can potentially be reconstructed as displaced from the primary $pp$ vertex (PV) resulting in $\effr \neq 1$; however, only the \ta dependence of the detection efficiency is required to use Eq.\,\eqref{eq:norm}.
Finally,  Eq.\,\eqref{eq:norm} is altered for large \ma to account for additional kinetic mixing with the $Z$ boson~\cite{Cassel:2009pu,Cline:2014dwa}.

The \lhcb detector is a single-arm forward spectrometer covering the pseudorapidity range $2<\eta <5$, described in detail in Refs.~\cite{Alves:2008zz,LHCb-DP-2014-002}.
The prompt-like \aprime search is based on a data sample that employs a novel data-storage strategy, made possible by advances in the LHCb data-taking scheme introduced in 2015~\cite{LHCb-PROC-2015-011,Aaij:2016rxn},
where all online-reconstructed particles are stored, but most lower-level information is discarded, greatly reducing the event size.
In contrast, the data sample used in the long-lived \aprime search is derived from the standard LHCb data stream.
Simulated data samples, which are used to validate the analysis, are produced using the software described in Refs.~\cite{Sjostrand:2014zea,*Sjostrand:2007gs,LHCb-PROC-2010-056,Allison:2006ve, *Agostinelli:2002hh}.

The online event selection is performed by a trigger~\cite{LHCb-DP-2012-004} consisting of a hardware stage using information from the calorimeter and muon systems, followed by a software stage that performs a full event
reconstruction.
At the hardware stage, events are required to have a muon with  momentum transverse to the beam direction ${\pt(\mu) \gtrsim 1.8\gev}$,
or a dimuon pair with $\pt(\mu^+) \pt(\mu^-) \gtrsim (1.5\gev)^2$.
%The long-lived \aprime search also uses events selected at the hardware stage independently of the \atomm candidate.
The long-lived \aprime search also uses events selected at the hardware stage due to the presence of a high-\pt hadron that is not associated to the \atomm candidate.
In the software stage, where the \pt resolution is substantially improved {\em cf.}\ the hardware stage,
\atomm candidates are built from two oppositely charged tracks
that form a good-quality vertex and satisfy stringent muon-identification criteria, though these criteria were loosened considerably in the low-mass region during 2017--2018 data taking.
Both searches require $\pt(\aprime) > 1\gev$ and $2<\eta(\mu)<4.5$.
The prompt-like \aprime search uses muons that are consistent with originating from the PV,
with ${\pt(\mu) > 1.0\gev}$ and momentum ${p(\mu) > 20\gev}$ in 2016,  and $\pt(\mu) > 0.5\gev$, ${p(\mu) > 10\gev}$, and ${\pt(\mu^+) \pt(\mu^-) > (1.0\gev)^2}$ in 2017--2018.
The long-lived \aprime search uses muons that are inconsistent with originating from any PV with $\pt(\mu) > 0.5\gev$ and ${p(\mu) > 10\gev}$,
and requires $2 < \eta(\aprime)<4.5$ and a decay topology consistent with a dark photon originating from a PV.

The prompt-like \aprime sample is contaminated by
prompt ${\gtomm}$ production,
various resonant decays to $\mu^+\mu^-$, whose mass-peak regions are avoided in the search,
and by the following types of misreconstruction:
($hh$) two prompt hadrons misidentified as muons;
($h\mu_Q$) a misidentified prompt hadron combined with a muon produced in the decay of a heavy-flavor quark, $Q$, that is misidentified as prompt;
and
($\mu_Q\mu_Q$) two muons produced in $Q$-hadron decays that are both  misidentified as prompt.
Contamination from a prompt muon and a misidentified prompt hadron is negligible, though it is accounted for automatically by the method used to determine the sum of the $hh$ and $h\mu_Q$ contributions.
The impact of the \gtomm background is reduced, {\em cf.}\ Ref.~\cite{LHCb-PAPER-2017-038}, by constraining the muons to originate from the PV when determining \mmm.
This improves the resolution, \sa, by about a factor of $2$ for small \ma.
The misreconstructed backgrounds are highly suppressed by the stringent %muon-identification and prompt-like
requirements applied in the trigger; however, substantial contributions remain for $\ma \gtrsim 1.1\gev$.
In this mass region, dark photons are expected to be predominantly produced in Drell--Yan processes,
from which they would inherit the well-known signature of dimuon pairs that are largely isolated.
Therefore, the signal sensitivity is enhanced by applying the anti-$k_{\rm T}$-based~\cite{antikt,fastjet,LHCb-PAPER-2013-058} isolation requirement described in Refs.~\cite{LHCb-PAPER-2017-038,Supp} for $\ma > 1.1\gev$.

\begin{figure}[t!]
  \centering
  \includegraphics[width=0.95\textwidth]{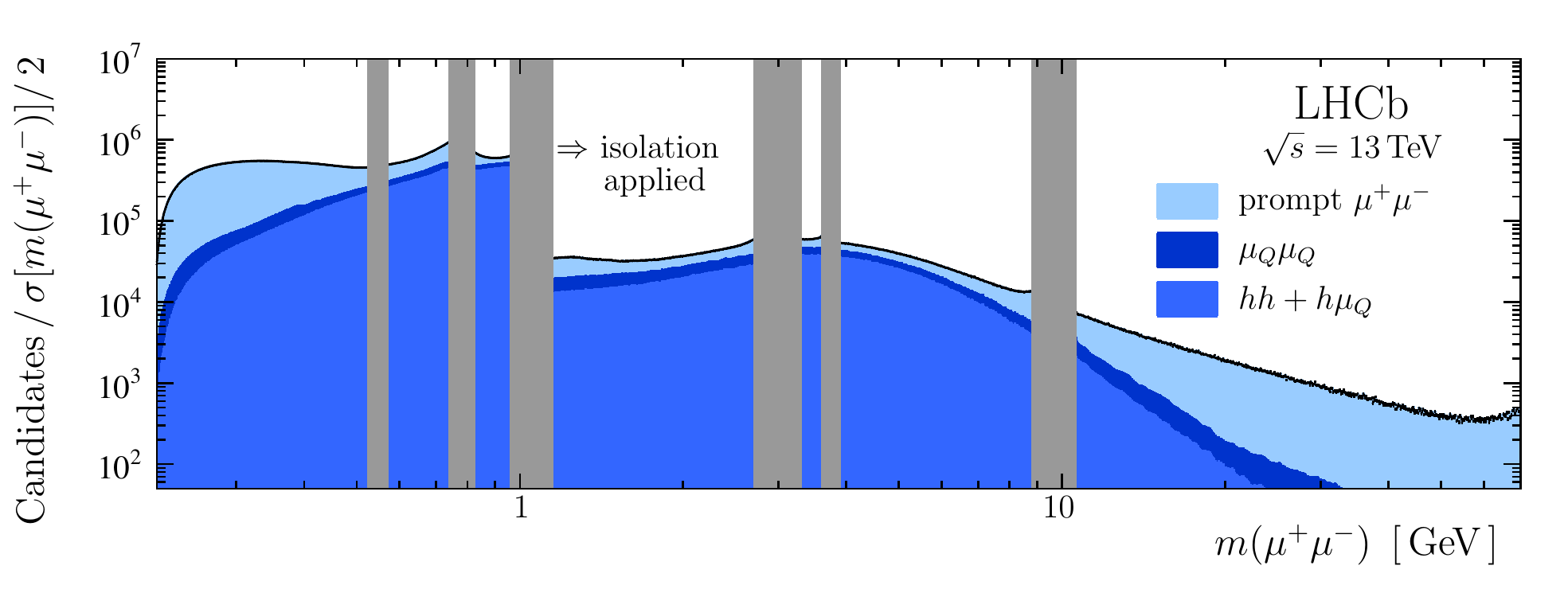}
  \caption{
  Prompt-like mass spectrum,
  where the categorization of the data as prompt $\mu^+\mu^-$, $\mu_Q\mu_Q$, and $hh+h\mu_Q$ is determined using the \mxip fits described in the text (examples of these fits are provided in the Supplemental Material).
  The anti-$k_{\rm T}$-based isolation requirement is applied for $\ma > 1.1\gev$.
  }
  \label{fig:prompt_bytype}
\end{figure}

The observed prompt-like \atomm yields, which are determined from fits to the \mmm spectrum, are normalized using Eq.\,\eqref{eq:norm} to obtain constraints on $\varepsilon^2$.
The \ngob values in Eq.\,\eqref{eq:norm} are obtained from binned extended maximum likelihood fits to the \mxip distributions, where \xip is defined as the difference in the vertex-fit $\chi^2$ when the PV is reconstructed with and without the muon. The \mxip distribution provides excellent discrimination between prompt muons and the displaced muons that constitute the $\mu_Q\mu_Q$ background.
%Since \xip approximately follows a \chisq probability density function (PDF), with two degrees of freedom, the \mxip distributions have minimal mass dependence for each source of dimuon candidates.
The \xip quantity approximately follows a \chisq probability density function (PDF), with two degrees of freedom, and therefore, the \mxip distributions have minimal dependence on mass for each source of dimuon candidates.
The prompt-dimuon PDFs are taken directly from data at $m(\jpsi)$ and $m(Z)$, where prompt resonances are dominant.
Small
corrections are applied to obtain these PDFs at all other \ma, which are validated near threshold, at $m(\phi)$, and at $m[\Upsilon(1S)]$, where the data predominantly consist of prompt dimuon pairs.
Based on these validation studies, a shape uncertainty of 2\% is applied in each \mxip bin.
Same-sign $\mu^{\pm}\mu^{\pm}$ candidates provide estimates for the PDF and yield of the sum of the $hh$ and $h\mu_Q$ contributions, where each involves misidentified prompt hadrons.
The $\mu^{\pm}\mu^{\pm}$ yields are corrected to account for the difference in the production rates of $\pi^{+}\pi^{-}$ and $\pi^{\pm}\pi^{\pm}$, which are determined precisely from data using dipion candidates weighted to account for the kinematic dependence of the muon  misidentification probability,
since the $hh$ background largely consists of
 $\pi^{+}\pi^{-}$ pairs where both pions are misidentified.
The uncertainty due to the finite size of the $\mu^{\pm}\mu^{\pm}$ sample in each bin is included in the likelihood.
Simulated $Q$-hadron decays are used to obtain the $\mu_Q\mu_Q$ PDFs, where the dominant uncertainties are from the relative importance of the various $Q$-hadron decay contributions at each mass.
Example \mxip fits are provided in Ref.~\cite{Supp}, while the resulting prompt-like candidate categorization versus \mmm is shown in Fig.~\ref{fig:prompt_bytype}.
Finally, the \ngob yields are corrected for bin migration due to bremsstrahlung, which is negligible except near the low-mass tails of the \jpsi and $\Upsilon(1S)$, and the small expected Bethe--Heitler contribution is subtracted~\cite{Ilten:2016tkc}, resulting in the \naex values shown in %Fig.~\ref{fig:prompt_nexp}
Fig.~S2
of Ref.~\cite{Supp}.

The prompt-like \naob mass spectrum is scanned in steps of \sa/2 searching for \atomm contributions\,\cite{Supp}, using the  strategy from  Ref.~\cite{LHCb-PAPER-2017-038}.
At each mass, a binned extended maximum likelihood fit is performed in a $\pm 12.5\,\sa$ window around \ma.
The profile likelihood is used to determine the $p$-value and the upper limit at  90\% confidence level (CL) on \naob.
The signal is well modeled by a Gaussian distribution whose resolution is determined with 10\% precision using a combination of simulated \atomm decays and the observed \pt-dependent widths of the large resonance peaks in the data. The mass-resolution uncertainty is included in the profile likelihood.
The method of Ref.~\cite{Williams:2017gwf} selects the background model from a large set of potential components, which includes all Legendre modes up to tenth order and dedicated terms for known resonances, by performing a data-driven process whose uncertainty is included in the profile likelihood following Ref.~\cite{Dauncey:2014xga}.
No significant excess is found in the prompt-like \ma spectrum, after accounting for the trials factor due to the number of signal hypotheses.

\begin{figure}
  \centering
  \includegraphics[width=0.99\textwidth]{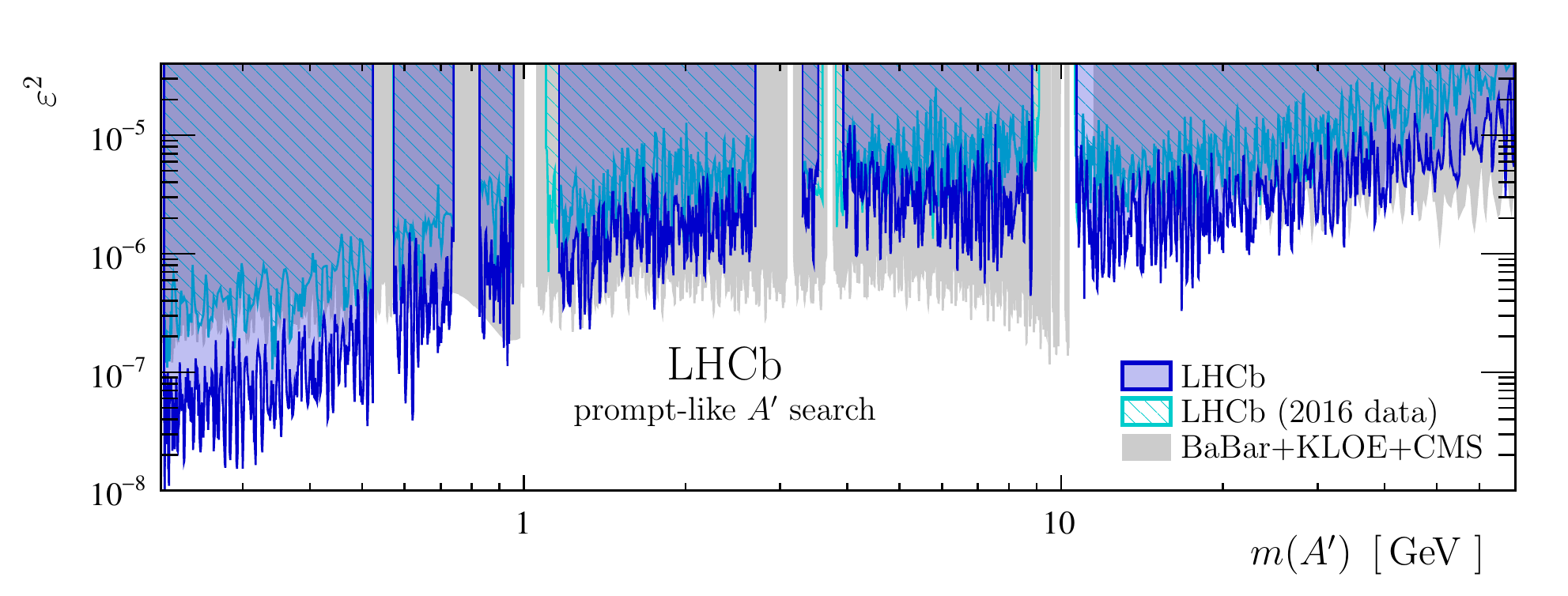}
  \caption{
  Regions of the \maeps parameter space excluded at 90\% CL by the  prompt-like \aprime search compared to the best published~\cite{Lees:2014xha,Anastasi:2018azp,LHCb-PAPER-2017-038} and preliminary~\cite{CMS-PAS-EXO-19-018} limits.
  }
  \label{fig:lims_prompt}
\end{figure}

Dark photons are excluded at 90\% CL where the upper limit on \naob is less than \naex.
Figure~\ref{fig:lims_prompt} shows that the constraints placed on prompt-like dark photons are the most stringent for $214 < \ma \lesssim 740\mev$ and $10.6 < \ma \lesssim 30\gev$.
The low-mass constraints are the strongest placed by a prompt-like \aprime search at any \ma.
These results are corrected for inefficiency and changes in the mass resolution that arise due to \ta no longer being negligible at such small values of $\epsilon^2$.
The high-mass constraints are adjusted to account for additional kinetic mixing with the $Z$ boson~\cite{Cassel:2009pu,Cline:2014dwa}, which alters Eq.\,\eqref{eq:norm}.
Since the LHCb detector response is independent of which $q\bar{q}\to\aprime$ process produces the dark photon above 10\gev, it is straightforward to recast the results in Fig.~\ref{fig:lims_prompt} for other models~\cite{Ilten:2018crw,Fayet:1990wx}.

For the long-lived \aprime search, contamination from prompt particles is negligible due to a stringent criterion applied in the trigger on \mxip that requires muons be inconsistent with originating from any PV.
Therefore, the dominant background contributions are:
photons that convert into $\mu^+\mu^-$ in the silicon-strip vertex detector that surrounds the $pp$ interaction region, known as the VELO~\cite{LHCb-DP-2014-001};
$b$-hadron decay chains that produce two muons;
and the low-mass tail from $\KS\to\pi^+\pi^-$ decays, where both pions are misidentified as muons (all other strange decays are negligible).
A $p$-value is assigned to the photon-conversion hypothesis for each long-lived \atomm candidate using properties of the decay vertex and muon tracks, along with a high-precision three-dimensional material map produced from a data sample of secondary hadronic interactions~\cite{LHCb-DP-2018-002}.
A \ma-dependent requirement is applied to these $p$-values that results in conversions having negligible impact on the sensitivity, though they are still accounted for to prevent pathologies when there are no other background sources.
The remaining backgrounds are highly suppressed by the decay topology requirement applied in the trigger.
Furthermore, since muons produced in $b$-hadron decays are often accompanied by additional displaced tracks,
events are rejected if they are selected by the inclusive heavy-flavor software trigger~\cite{Likhomanenko:2015aba,BBDT} independently of the presence of the \atomm candidate.
In addition, boosted decision tree classifiers are used to reject events containing tracks consistent with originating from the same $b$-hadron decay as the signal muon candidates~\cite{LHCb-PAPER-2017-001}.

\begin{figure}
  \centering
  \includegraphics[width=0.68\textwidth]{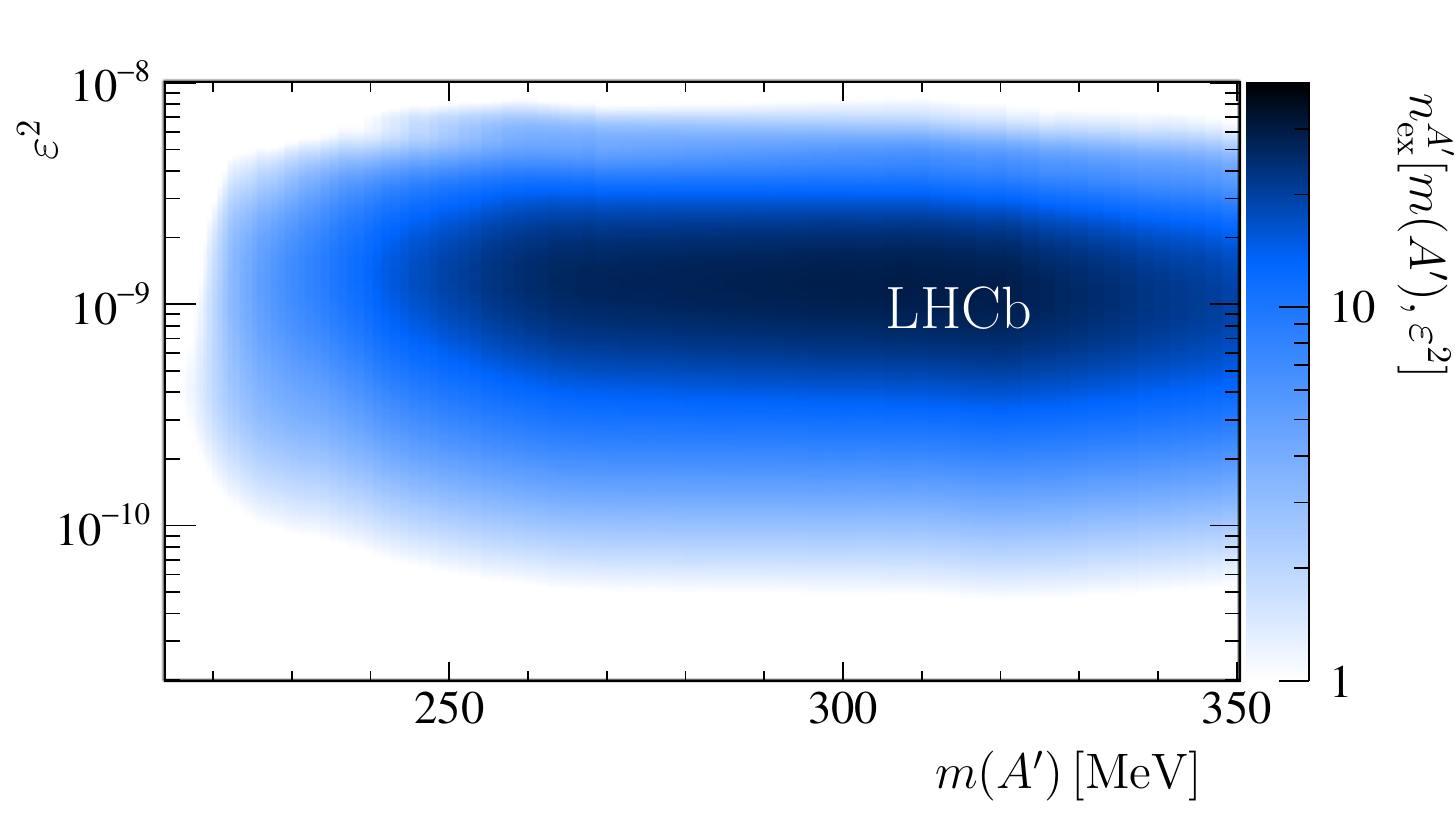}
  \caption{
  Expected reconstructed and selected long-lived \atomm yield.
  }
  \label{fig:displ_nexp}
\end{figure}

The long-lived \aprime search is also normalized using Eq.\,\eqref{eq:norm}; however, \effr is not unity, in part because the efficiency depends on the decay time, $t$.
The kinematics are identical for \atomm and prompt \gtomm decays for $\ma = m(\gamma^*)$;
therefore, the $t$ dependence of \effr is obtained by resampling prompt \gtomm candidates as long-lived \atomm decays, where all $t$-dependent properties, {\em e.g.}\ \mxip,
are recalculated based on the resampled decay-vertex locations (the impact of background contamination in the prompt \gtomm sample is negligible).
This approach is validated using simulation, where prompt \atomm decays are used to predict the properties of long-lived \atomm decays.
The relative uncertainty on \effr is estimated to be 5\%, which arises largely due to limited knowledge of how radiation damage affects the performance of the VELO as a function of the distance from the $pp$ interaction region.
The looser kinematic, muon-identification, and hardware-trigger requirements applied to long-lived \atomm candidates, {\em cf.}\ prompt-like candidates, also increase the efficiency.
This $t$-independent increase in efficiency is determined using a  control data sample of dimuon candidates consistent with originating from the PV, but otherwise satisfying the long-lived criteria.
The \naex values obtained using these data-driven \effr values (discussed in more detail in Ref.~\cite{Supp}), along with the expected prompt-like \atomm yields, %in Fig.~\ref{fig:prompt_nexp},
are shown in Fig.~\ref{fig:displ_nexp}.

The long-lived \ma spectrum is also scanned in discrete steps of \sa/2 looking for \atomm contributions\,\cite{Supp}; however, discrete steps in \ta are also considered here.
Binned extended maximum likelihood fits are performed to the three-dimensional feature space of \mmm, $t$,
and the consistency of the decay topology as quantified in the decay-fit $\chi^2_{\rm DF}$, which has three degrees of freedom.
The photon-conversion contribution is derived in each $[\mmm, t, \chi^2_{\rm DF}]$ bin from the number of dimuon candidates that are rejected by the conversion criterion.
Both the $b$-hadron and \KS contributions are modeled in each $[t, \chi^2_{\rm DF}]$ bin by second-order polynomials of the energy released in the decay,
$\sqrt{\mmm^2 - 4 m(\mu)^2}$.
These contributions are validated using the following large control data samples:
candidates that fail the $b$-hadron suppression requirements;
and
candidates that fail, but nearly satisfy, the stringent muon-identification requirements.
The profile likelihood is used to obtain the $p$-values and confidence intervals on \naobt.
No significant excess is observed in the long-lived \atomm search (the three-dimensional data distribution and the background-only pull distributions are provided in Ref.~\cite{Supp}).

\begin{figure}
  \centering
  \includegraphics[width=0.68\textwidth]{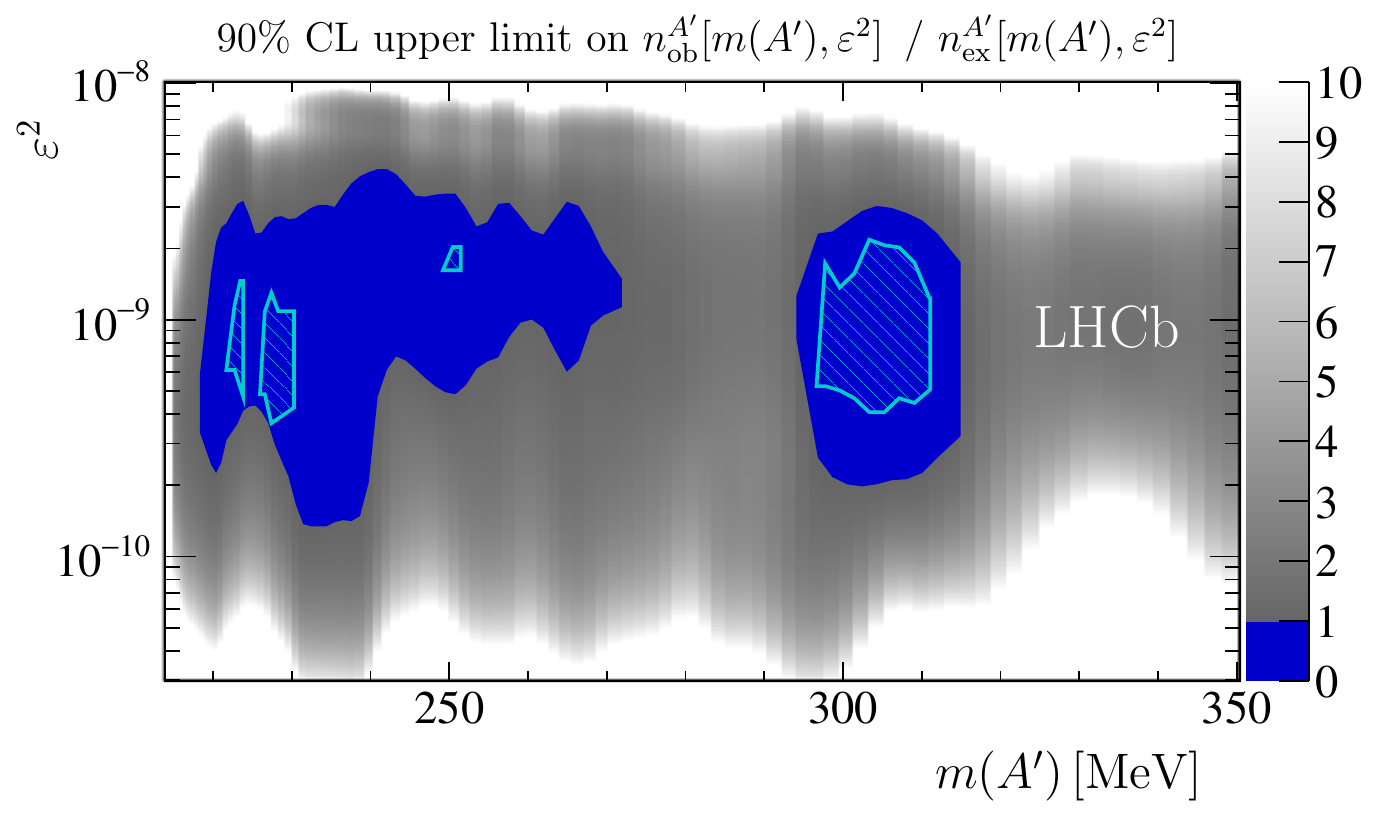}
  \caption{
  Ratio of the observed upper limit on \naobe at 90\% CL to the expected dark-photon yield, \naex, where regions less than unity are excluded. The only  constraints in this region are from (hashed) the previous LHCb search~\cite{LHCb-PAPER-2017-038}.
  }
  \label{fig:displ_lims}
\end{figure}

Since the relationship between \ta and $\varepsilon^2$ is known at each mass~\cite{Ilten:2016tkc}, the upper limits on \naobt are easily translated into limits on \naobe.
Regions of the \maeps parameter space where the upper limit on \naobe is less than \naex are excluded at 90\% CL.
Figure~\ref{fig:displ_lims} shows that sizable regions of $[\ma,\varepsilon^2]$ parameter space are excluded, which are much larger than those excluded in Ref.~\cite{LHCb-PAPER-2017-038}.
%Furthermore, most of the parameter space shown in Fig.~\ref{fig:displ_lims} would have been accessible if the  data sample was roughly three times larger.
%The expected number of recorded \atomm decays should increase by a factor $\mathcal{O}(100)$ in the
%data sample to be collected in Run~3 by the upgraded LHCb detector.

In summary, searches are performed for prompt-like and long-lived dark photons produced in $pp$ collisions at a center-of-mass energy of 13\tev.
Both searches look for \atomm decays using a data sample corresponding to an integrated luminosity of 5.5\invfb collected with the LHCb detector during 2016--2018.
No evidence for a signal is found in either search, and 90\% CL exclusion regions are set on the $\gamma$--$\aprime$ kinetic-mixing strength.
The prompt-like \aprime search is performed from near the dimuon threshold up to 70\gev, and produces the most stringent constraints on dark photons with $214  < \ma \lesssim 740\mev$ and $10.6 < \ma \lesssim 30\gev$.
The long-lived \aprime search
is restricted to the mass range $214<\ma<350\mev$, where the data sample potentially has sensitivity, and places world-leading constraints on low-mass dark photons with lifetimes $\mathcal{O}(1)\ps$.
The three-fold increase in integrated luminosity, improved trigger efficiency during 2017--2018 data taking, and improvements in the analysis result in the searches presented in this Letter achieving much better sensitivity to dark photons than the previous LHCb results~\cite{LHCb-PAPER-2017-038}.
The prompt-like \aprime search achieves a factor of 5\,(2) better sensitivity to $\varepsilon^2$ at low\,(high) masses than Ref.~\cite{LHCb-PAPER-2017-038},
while the long-lived \aprime search provides access to much larger regions of \maeps parameter space.

These results demonstrate the excellent sensitivity of the LHCb experiment to dark photons, even using a data sample collected with a hardware-trigger stage that is highly inefficient for low-mass \atomm decays.
%
%The expected number of recorded \atomm decays should increase by a factor $\mathcal{O}(100)$ in the data sample to be collected in Run~3 by the upgraded LHCb detector.
%
The removal of this hardware-trigger stage in Run~3, along with the planned increase in luminosity, should increase the potential yield of  \atomm decays in the low-mass region by a factor $\mathcal{O}(100)$ compared to the 2016--2018 data sample.
Given that most of the parameter space shown in Fig.~\ref{fig:displ_lims} would have been accessible if the data sample was only three times larger, these upgrades will greatly increase the dark-photon discovery potential of the LHCb experiment.

\section*{Acknowledgements}

\noindent We express our gratitude to our colleagues in the CERN
accelerator departments for the excellent performance of the LHC. We
thank the technical and administrative staff at the LHCb
institutes.
We acknowledge support from CERN and from the national agencies:
CAPES, CNPq, FAPERJ and FINEP (Brazil);
MOST and NSFC (China);
CNRS/IN2P3 (France);
BMBF, DFG and MPG (Germany);
INFN (Italy);
NWO (Netherlands);
MNiSW and NCN (Poland);
MEN/IFA (Romania);
MSHE (Russia);
MinECo (Spain);
SNSF and SER (Switzerland);
NASU (Ukraine);
STFC (United Kingdom);
DOE NP and NSF (USA).
We acknowledge the computing resources that are provided by CERN, IN2P3
(France), KIT and DESY (Germany), INFN (Italy), SURF (Netherlands),
PIC (Spain), GridPP (United Kingdom), RRCKI and Yandex
LLC (Russia), CSCS (Switzerland), IFIN-HH (Romania), CBPF (Brazil),
PL-GRID (Poland) and OSC (USA).
We are indebted to the communities behind the multiple open-source
software packages on which we depend.
Individual groups or members have received support from
AvH Foundation (Germany);
EPLANET, Marie Sk\l{}odowska-Curie Actions and ERC (European Union);
ANR, Labex P2IO and OCEVU, and R\'{e}gion Auvergne-Rh\^{o}ne-Alpes (France);
Key Research Program of Frontier Sciences of CAS, CAS PIFI, and the Thousand Talents Program (China);
RFBR, RSF and Yandex LLC (Russia);
GVA, XuntaGal and GENCAT (Spain);
the Royal Society
and the Leverhulme Trust (United Kingdom).

\setboolean{inbibliography}{true}
\bibliographystyle{LHCb}
\bibliography{dark-photons,main,standard,LHCb-PAPER,LHCb-CONF,LHCb-DP,LHCb-TDR}

\newpage
\clearpage
\newpage
\setcounter{equation}{0}
\setcounter{figure}{0}
\setcounter{table}{0}
\setcounter{section}{0}
\setcounter{page}{1}
\makeatletter
\renewcommand{\theequation}{S\arabic{equation}}
\renewcommand{\thefigure}{S\arabic{figure}}
\renewcommand{\thetable}{S\arabic{table}}
\renewcommand{\thepage}{S\arabic{page}}
\newcommand\ptwiddle[1]{\mathord{\mathop{#1}\limits^{\scriptscriptstyle(\sim)}}}

\begin{center}
\textbf{\Large Supplemental Material for LHCb-PAPER-2019-031} \\
\vspace{0.15in}
\textbf{\Large \it Search for \atomm decays} \\
\end{center}

Additional details and figures for the prompt-like and long-lived searches are presented in this Supplemental Material.

\vspace{0.2in}

\noindent \textbf{Isolation Criterion} \vspace{0.05in}

For masses above the $\phi(1020)$ meson mass, dark photons are expected to be predominantly produced in Drell--Yan processes in $pp$ collisions at the LHC.
A well-known signature of Drell--Yan production is dimuon pairs that are largely isolated, and a high-mass dark photon would inherit this property.
The signal sensitivity for $\ma > 1.1\gev$ is enhanced by applying the jet-based isolation requirement used in Ref.~\cite{LHCb-PAPER-2017-038}.
Jet reconstruction is performed by clustering charged and neutral particle-flow candidates~\cite{LHCb-PAPER-2013-058}
using the anti-$k_{\rm T}$ clustering algorithm~\cite{antikt} with $R=0.5$ as implemented in \textsc{FastJet}~\cite{fastjet}.
Muons with $\pt(\mu)/\pt({\rm jet}) < 0.7$ are rejected, where the contribution to $\pt({\rm jet})$ from the other muon is excluded if both muons are clustered in the same jet,
as this is found to provide nearly optimal sensitivity for all $\ma > m(\phi)$.

\vspace{0.2in}

\noindent \textbf{Prompt-Like Fits}\vspace{0.05in}

The prompt-like fit strategy is the same as described in detail in the Supplemental Material of Ref.~\cite{LHCb-PAPER-2017-038}.
This strategy was first introduced in Ref.~\cite{Williams:2017gwf}, where it is denoted by aic-o.
The \mmm spectrum is scanned in steps of \sa/2 searching for ${\atomm}$ contributions.
At each mass, a binned extended maximum likelihood fit is performed, and the profile likelihood is used to determine the $p$-value and the confidence interval on \naob.
The prompt-like-search trials factor is obtained using pseudoexperiments.
As in Ref.~\cite{Williams:2017gwf}, each fit is performed in a $\pm12.5\,\sa$ window around the scan-mass value using bins with widths of $\sa/20$.
Near threshold, the energy released in the decay, $\sqrt{\mmm^2 - 4m(\mu)^2}$, is used instead of the mass since it is easier to model.
The confidence intervals are defined using the {\em bounded likelihood} approach,
 which involves taking $\Delta \log{\mathcal{L}}$ relative to zero signal, rather than the best-fit value, if the best-fit signal value is negative.
This approach enforces that only physical (nonnegative) upper limits are placed on \naob, and prevents defining exclusion regions that are much better than the experimental sensitivity in cases where a large deficit in the background yield is observed.

The signal models are determined at each \ma using a combination of simulated \atomm decays and the widths of the large resonance peaks that are clearly visible in the data.
The background models take as input a large set of potential background components, then the data-driven model-selection process of Ref.~\cite{Williams:2017gwf} is performed, whose uncertainty is included in the profile likelihood following Ref.~\cite{Dauncey:2014xga}.
Specifically, the method labeled {\em aic-o} in Ref.~\cite{Williams:2017gwf} is used, where the log-likelihood of each background model is penalized for its complexity (number of parameters).
The confidence intervals are obtained from the profile likelihoods, including the penalty terms, where the model index is treated as a discrete nuisance parameter, as originally proposed in Ref.~\cite{Dauncey:2014xga}. %using the Akaike Information Criterion(AIC)
In this analysis, the set of possible background components includes all Legendre modes with $\ell \leq 10$ at every \ma.
Additionally, dedicated background components are included for sizable narrow SM resonance contributions.
The use of 11 Legendre modes adequately describes every double-misidentified peaking background that contributes at a significant level, and therefore, these do not require dedicated background components.
In mass regions where such complexity is not required, the data-driven model-selection procedure reduces the complexity which increases the sensitivity to a potential signal contribution.
Therefore, the impact of the background-model uncertainty on the size of the confidence intervals is mass dependent, though on average it is about 30\%.
As in Ref.~\cite{Williams:2017gwf}, all fit regions are transformed onto the interval $[-1,1]$, where the scan \ma value maps to zero.
After such a transformation, the signal model is (approximately) an even function;
therefore, odd Legendre modes are orthogonal to the signal component, which means that the presence of odd modes has minimal impact on the variance of \naob.
In the prompt-like fits, all odd Legendre modes up to ninth order are included in every background model, while only a subset of the even modes is selected for inclusion in each fit.
%and the method labeled {\em aic-o} in Ref.~\cite{Williams:2017gwf} is used.
%Here, each background model contains every odd mode with $\ell \leq 9$, and all possible subsets of even modes with $\ell \leq 10$ are considered ($2^6 = 64$ total models).

Regions in the mass spectrum where large known resonance contributions are observed are vetoed in the prompt-like \aprime search.
Furthermore, the regions near the $\eta^{\prime}$ meson and the excited $\Upsilon$ states (beyond the $\Upsilon(3S)$ meson) are treated specially.
For example, since it is not possible to distinguish between \atomm and $\eta^{\prime}\!\to\!\mu^+\mu^-$ contributions at $m(\eta^{\prime})$, the $p$-values near this mass are ignored.
The small excess at $m(\eta^{\prime})$ is treated as signal when setting the limits on \naob, which is conservative in that an $\eta^{\prime}\!\to\!\mu^+\mu^-$ contribution will weaken the constraints on \atomm decays.
The same strategy is used near the excited $\Upsilon$ masses.

\vspace{0.2in}

\noindent \textbf{Long-Lived Fits}\vspace{0.05in}

\begin{figure}[h!]
  \centering
  \includegraphics[width=0.7\textwidth]{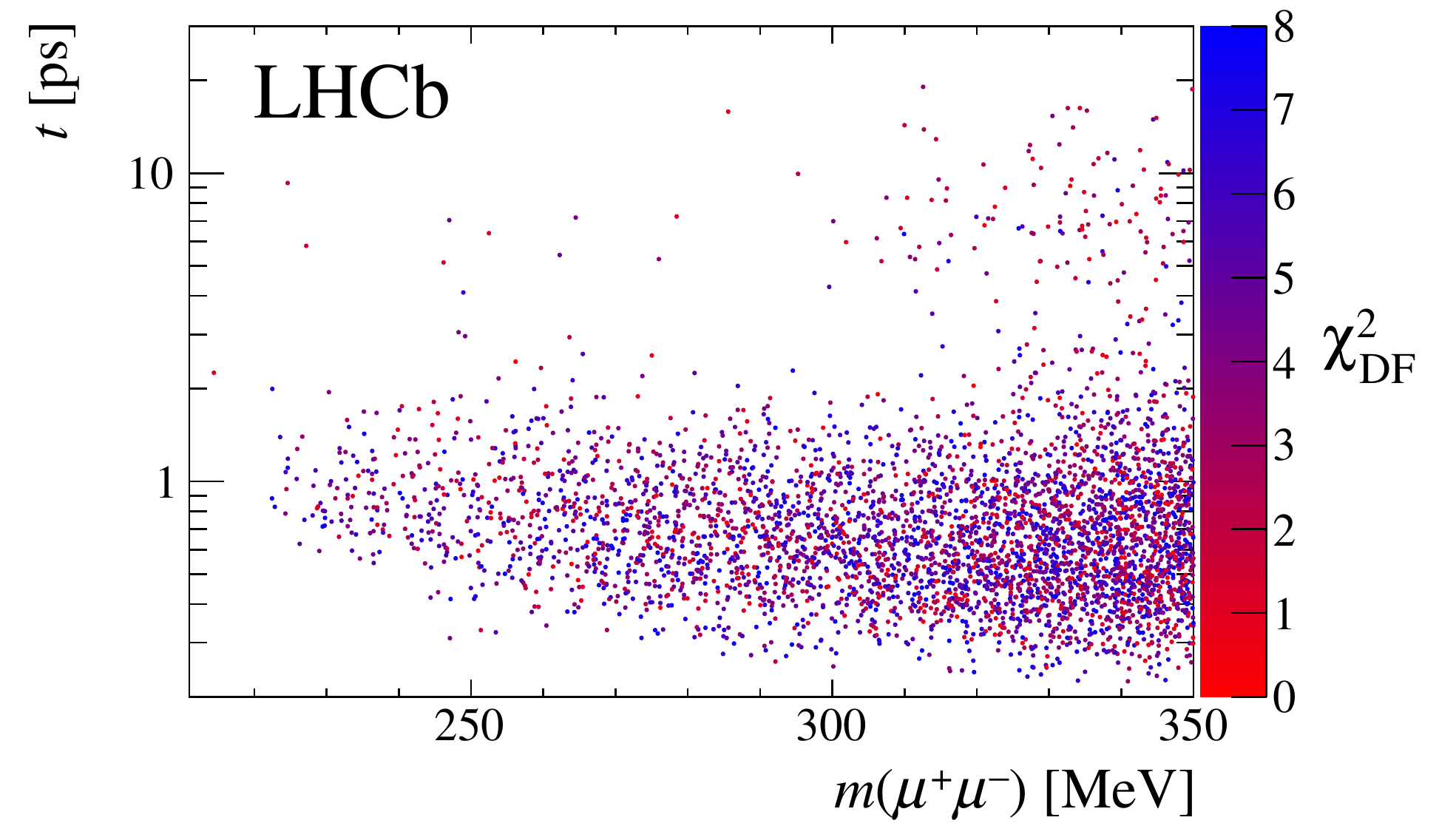}
  \caption{
  Three-dimensional distribution of $\chi^2_{\rm DF}$ versus $t$ versus \mmm, which is fit to determine the long-lived signal yields.
  The data are consistent with being predominantly due to $b$-hadron decays at small $t$, and due to \KS decays for large $t$ and $\mmm \gtrsim 280\mev$.
  The remaining few candidates at large $t$ and small \mmm are likely photon conversions.
  }
  \label{fig:displ_data}
\end{figure}

The long-lived fit strategy is also the same as described in the Supplemental Material to Ref.~\cite{LHCb-PAPER-2017-038}.
The signal yields are determined from binned extended maximum likelihood fits performed on all long-lived \atomm candidates using the three-dimensional feature space of \mmm, $t$, and $\chi^2_{\rm DF}$.
As in the prompt-like \aprime search, a scan is performed in discrete steps of \sa/2; however, in this case, discrete steps in \ta are also considered.
The profile likelihood is again used to obtain the $p$-values and the confidence intervals on \naobt.
The binning scheme involves four bins in $\chi^2_{\rm DF}$: [0,2], [2,4], [4,6], and [6,8].
Eight bins in $t$ are used: [0.2,0.6], [0.6,1.1], [1.1,1.6], [1.6,2.2], [2.2,3.0], [3,5], [5,10], and $>\!10\ps$.
The binning scheme used for \mmm depends on the scan \ma value, and is chosen such that the vast majority of the signal falls into a single bin.
Signal decays mostly have small $\chi^2_{\rm DF}$ values, with about 50\%\,(80\%) of \atomm decays satisfying $\chi^2_{\rm DF} < 2\,(4)$.
Background from $b$-hadron decays populates the small $t$ region and is roughly uniformly distributed in $\chi^2_{\rm DF}$, whereas background from \KS decays is signal-like in $\chi^2_{\rm DF}$ and roughly uniformly distributed in $t$.
Figure~\ref{fig:displ_data} shows the three-dimensional distribution of all long-lived \atomm candidates.

The expected contribution in each bin from photon conversions is derived from the number of candidates rejected by the conversion criterion.
Two large control data samples are used to develop and validate the modeling of the $b$-hadron and \KS contributions.
Both contributions are well modeled by second-order polynomials in $\sqrt{\mmm^2 - 4m(\mu)^2}$.
While no evidence for $t$ or $\chi^2_{\rm DF}$ dependence is observed for these parameters in either the $b$-hadron or \KS control sample, all parameters are allowed to vary independently in each $[t,\chi^2_{\rm DF}]$ region in the fits used in the long-lived \aprime search.

\begin{figure}[h!]
  \centering
  \includegraphics[width=0.49\textwidth]{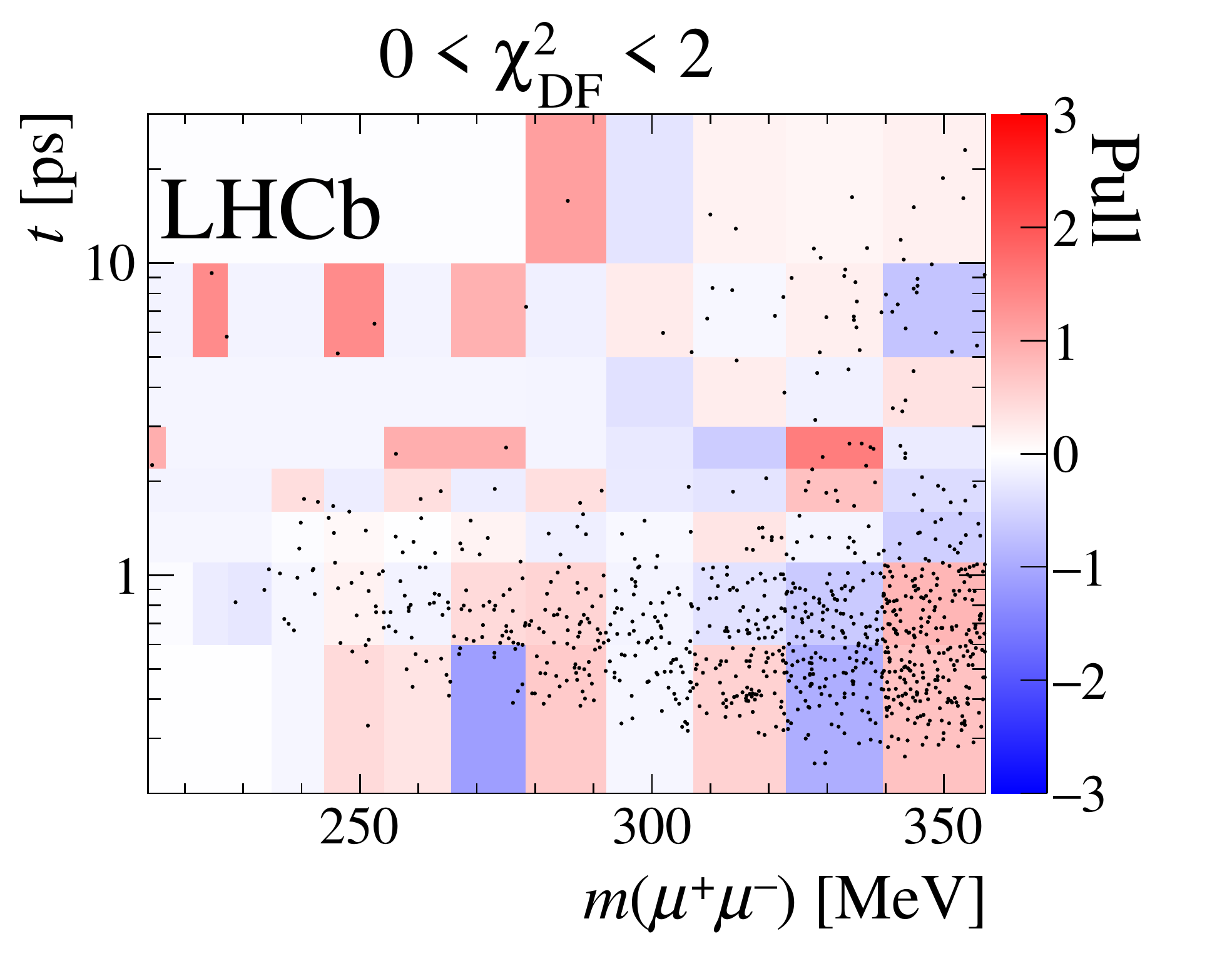}
  \includegraphics[width=0.49\textwidth]{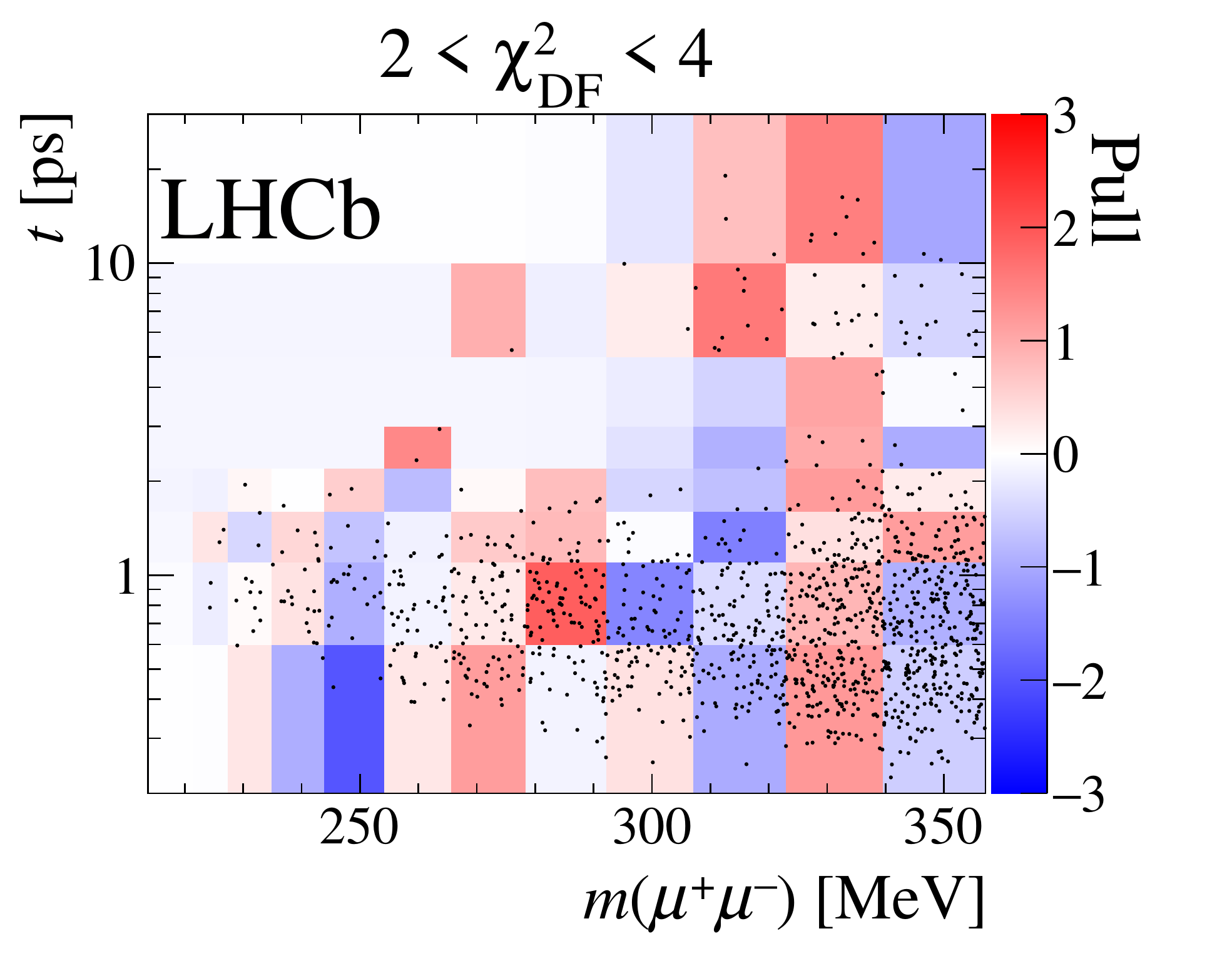}
  \includegraphics[width=0.49\textwidth]{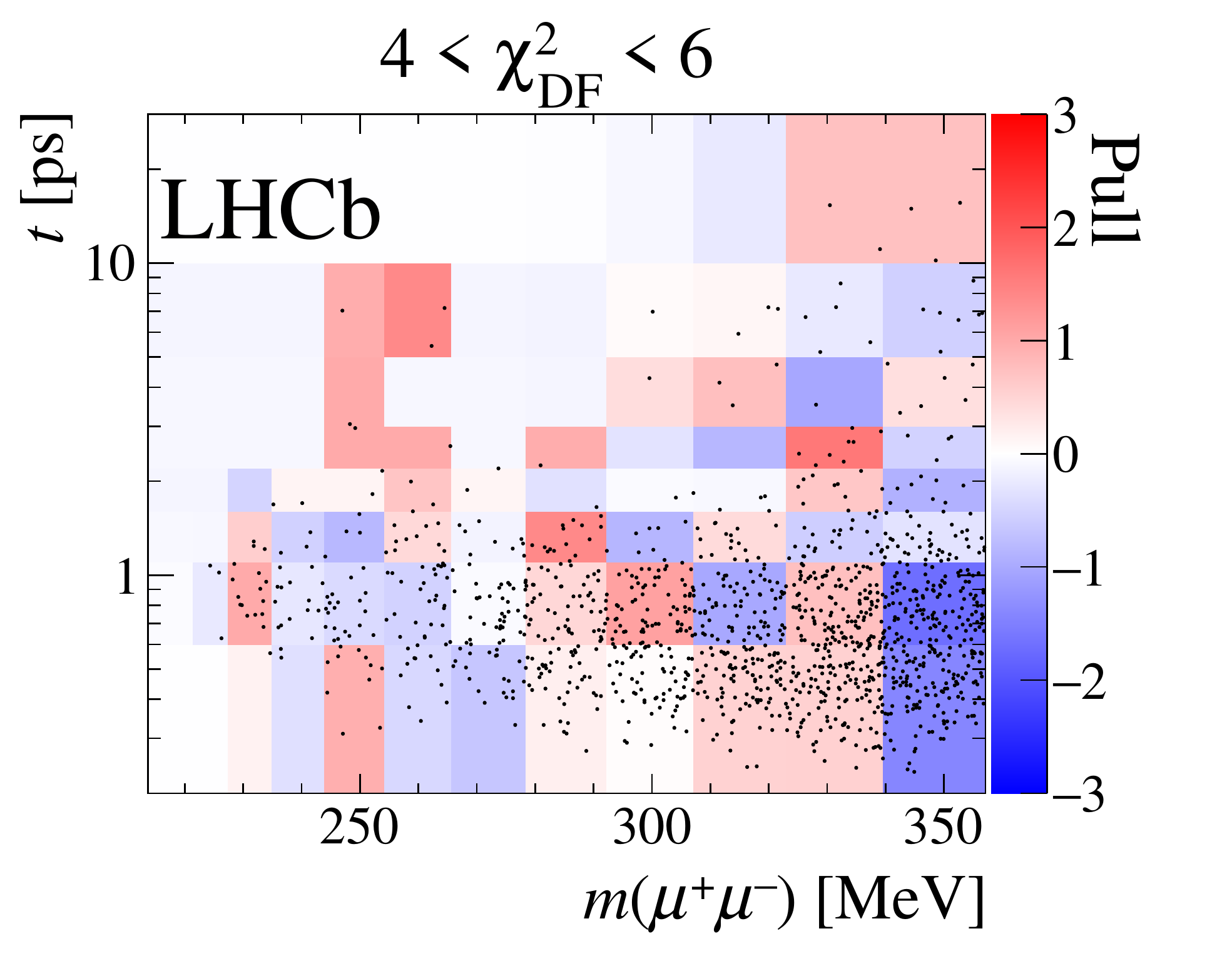}
  \includegraphics[width=0.49\textwidth]{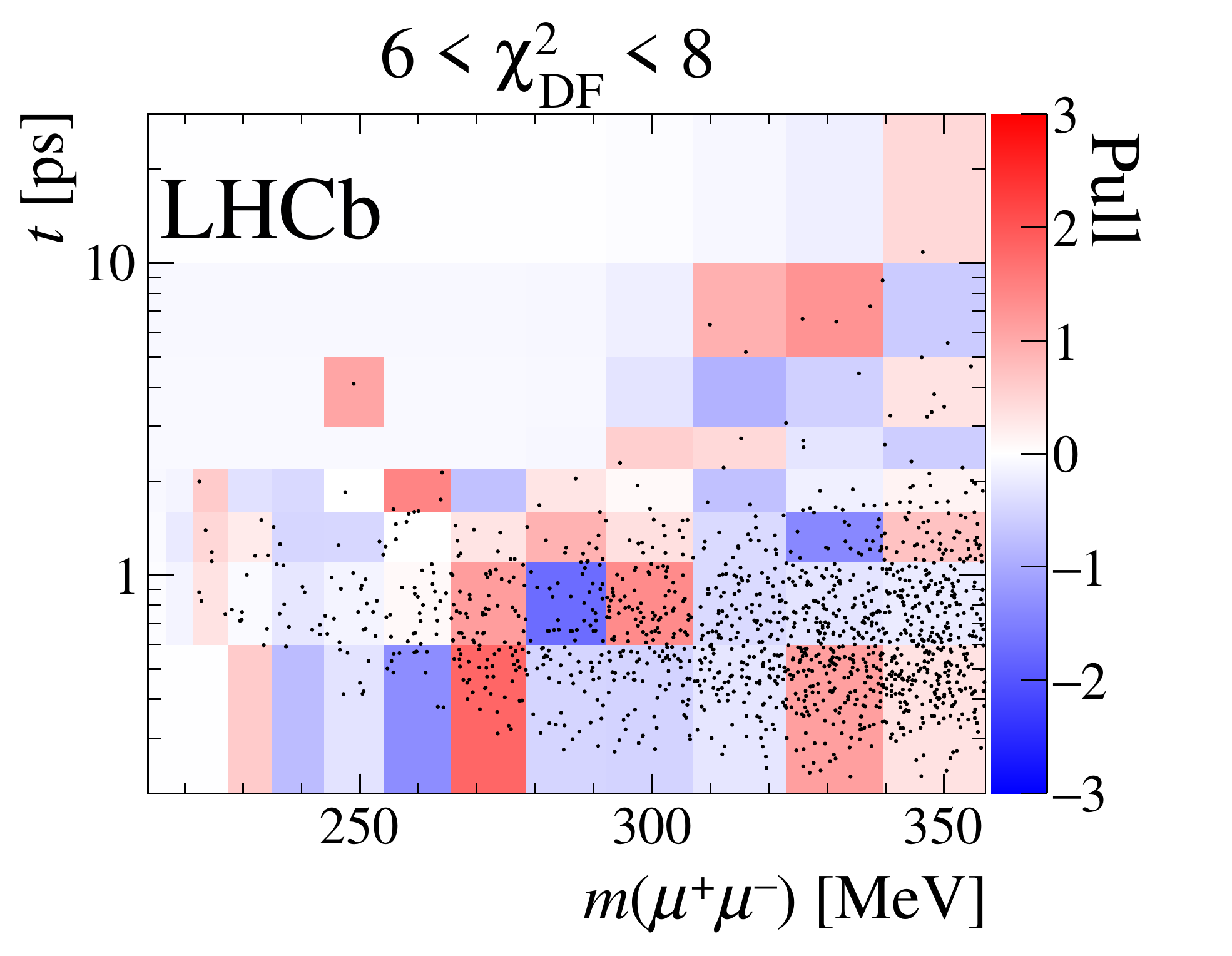}
  \caption{
  Long-lived \atomm candidates (black points) showing $t$ versus \mmm in bins of $\chi^2_{\rm DF}$, compared to the pulls from binned fits performed without a signal contribution (color axis).
  Positive pulls denote an excess of data candidates.
  }
  \label{fig:displ_pulls}
\end{figure}

Figure~\ref{fig:displ_pulls} shows the long-lived \atomm candidates, along with the pull values obtained from fits performed to the data where no signal contributions are included.
These distributions are consistent with those observed in pseudoexperiments where no signal component is generated.
Specifically, the data are consistent with being predominantly due to $b$-hadron decays at small $t$, and due to \KS decays for large $t$ and $\mmm \gtrsim 280\mev$.
The remaining few candidates at large $t$ and small \mmm are likely photon conversions.
{\em N.b.}, the same dominant backgrounds are expected in the upgraded LHCb detector in each region of mass and decay time.

\vspace{0.2in}

\noindent \textbf{Long-Lived Efficiency Determination}\vspace{0.05in}

The value of \effr is not unity in the long-lived \aprime search, in part because the efficiency depends on the decay time, $t$.
The kinematics are identical for \atomm and prompt \gtomm decays for $\ma = m(\gamma^*)$;
therefore, the $t$ dependence of \effr is obtained by resampling prompt \gtomm candidates as long-lived \atomm decays, where all $t$-dependent properties, {\em e.g.}\ \mxip,
are recalculated based on the resampled decay-vertex locations.
The impact of background contamination in the prompt \gtomm sample is shown to be negligible by recomputing \effr using same-sign $\mu^{\pm}\mu^{\pm}$ candidates, which are purely background, instead of prompt \gtomm candidates in this procedure.
The high-precision three-dimensional material map produced from a data sample of secondary hadronic interactions in Ref.~\cite{LHCb-DP-2018-002}, which forms the basis of the photon-conversion criterion applied in the selection, is used here to determine where each muon would hit active sensors, and thus, have recorded hits in the VELO.
The resolution on the vertex location and other $t$-dependent properties varies strongly with the location of the first VELO hit on each muon track, though this dependence is largely geometric, making rescaling the resolution of prompt tracks straightforward.
Finally, the values used for the decay-fit $\chi^2_{\rm DF}$ are resampled from the observed distribution for \KS decays in data, which is similar to that expected for an ideal $\chi^2$ distribution with three degrees of freedom, to ensure that non-Gaussian effects are properly considered.

\begin{figure}
  \centering
  \includegraphics[width=0.7\textwidth]{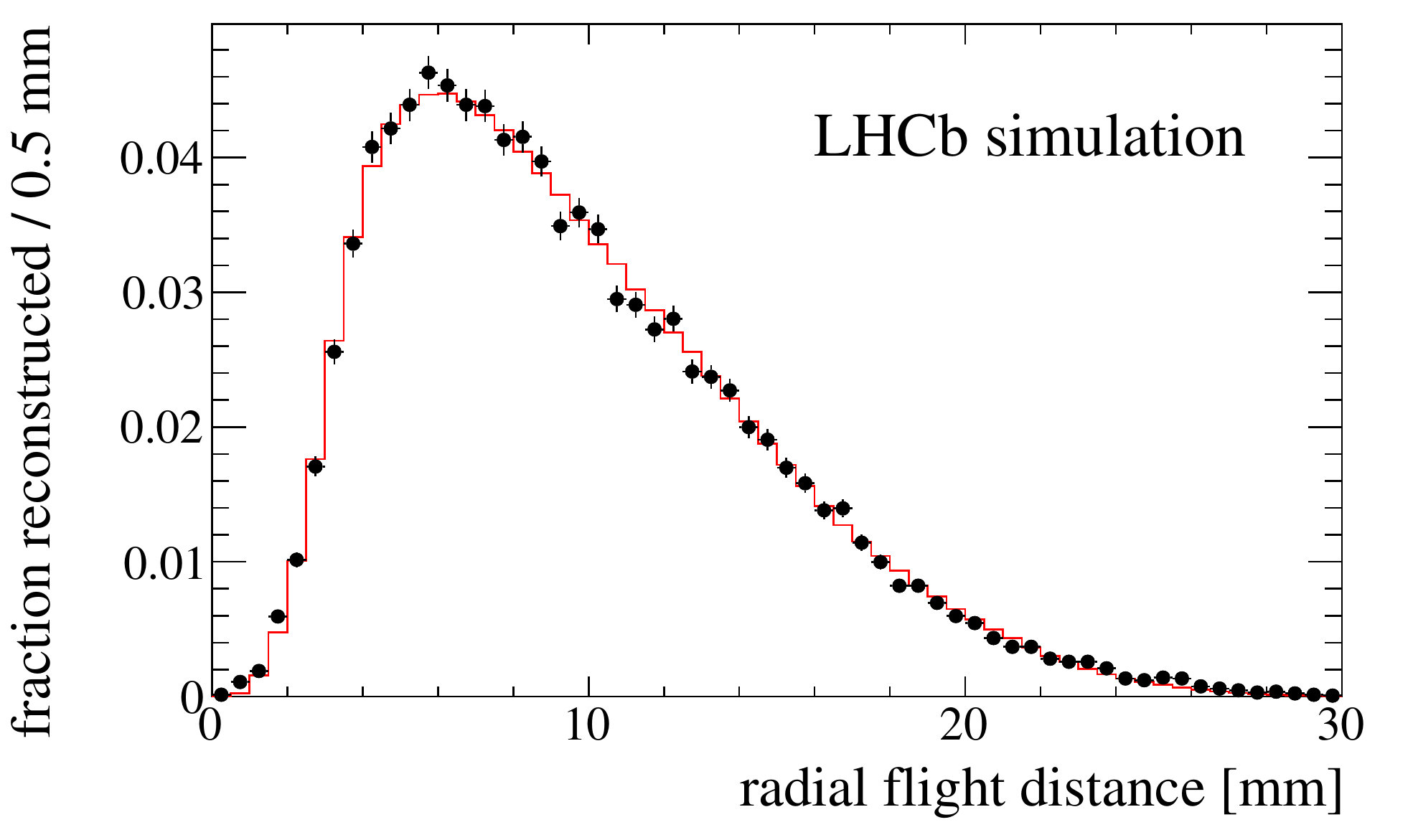}
  \caption{
  Validation of the data-driven approach used to predict the long-lived \aprime properties from prompt \gtomm candidates in data is shown using two
   simulated \atomm samples, each generated with the same production kinematics, for (black points) $\ta=2\ps$ reconstructed and satisfying the long-lived selection criteria except the photon-conversion criterion, and (red line) the prediction for $\ta=2\ps$ based only on the prompt \atomm sample. {\em N.b.}, there is no free normalization factor,  the red yield is predicted entirely from the prompt sample.
  }
  \label{fig:displ_sig_valid}
\end{figure}

This approach is validated using simulation, where prompt \atomm decays are used to predict the properties of long-lived \atomm decays.
Figure~\ref{fig:displ_sig_valid} shows one such example using \aprime decays generated with zero lifetime to predict the radial flight distance of a sample produced with $\ta = 2\ps$.
Here, the full long-lived selection is applied except for the photon-conversion criterion, which is validated separately as discussed below.
Since this validation study involves only simulated decays, the material map produced in Ref.~\cite{LHCb-DP-2018-002} is replaced by the geometry used in the simulation.
The integrated long-lived \atomm yield predictions agree to $\approx 2\%$ with the actual values in the $[\ma,\ta]$ region explored in the long-lived \aprime search.
Both the short-distance turn-on (driven by the \mxip criterion) and long-distance turn-off (driven by the minimum number of VELO hits required to form a track) curves are well described.
The dominant uncertainty here arises due to limited knowledge of how radiation damage has affected the performance of the VELO as a function of the distance from the $pp$ interaction region.
This uncertainty is estimated to be 5\% by rerunning the resampling method many times under different radiation-damage hypotheses.

To validate the efficiency of the photon-conversion criterion, the predicted efficiency for a dark photon with the same mass and lifetime as the \KS meson is compared to the efficiency observed in a control data sample of \KS decays.
The predicted and observed efficiencies agree to 1\%.
Additionally, in Ref.~\cite{LHCb-DP-2018-002} the expected performance of the photon-conversion criterion was shown to agree with the performance observed in an independent control data sample of photon conversions out to the $\mathcal{O}(10^{-4})$ level.
The expected performance here was obtained in a similar manner to how the dark photon efficiencies are obtained, except starting from a pure sample of conversions instead of prompt \gtomm candidates.

The looser kinematic, muon-identification, and hardware-trigger requirements applied to long-lived \atomm candidates, {\em cf.}\ prompt-like candidates, also increase the efficiency.
This $t$-independent increase in efficiency is determined using a  control data sample of dimuon candidates consistent with originating from the PV, but otherwise satisfying the long-lived criteria.
Specifically, \mxip fits like those used to determine \ngob in the prompt-like \aprime search are run on this control sample and on the subset of its candidates that satisfy the more stringent criteria applied in the prompt-like \aprime search.
The decrease in the \gtomm yield observed when applying the more stringent criteria at each mass is used to determine the $t$-independent increase in efficiency.

%\vspace{0.2in}
\clearpage

\noindent \textbf{Additional Figures}\vspace{0.05in}

This section provides additional figures from the analysis, and more detailed comparisons of the new constraints presented in the Letter with previous results.

\vspace{0.5in}

\begin{figure}[h!]
  \centering
  \includegraphics[width=0.95\textwidth]{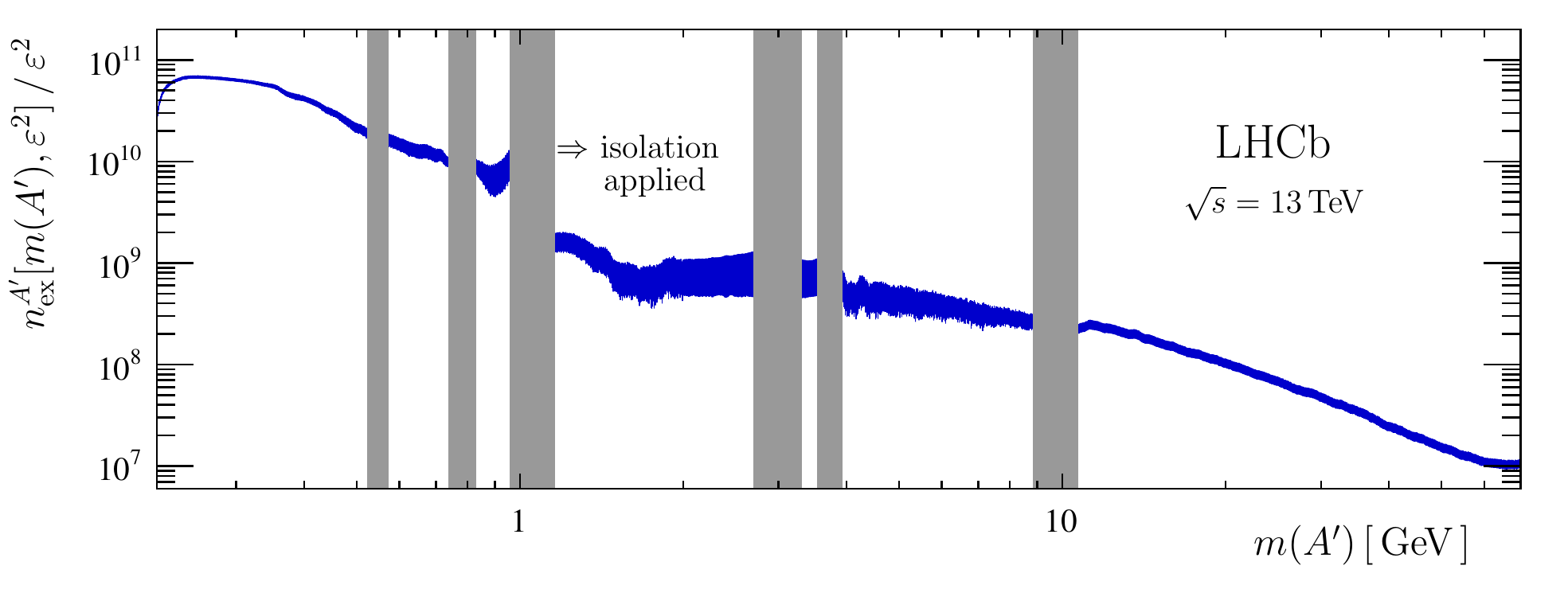}
  \caption{
  Expected reconstructed and selected prompt-like \atomm yield divided by $\varepsilon^2$, where the displayed uncertainties include the systematic contributions. The gray boxes cover the regions with large SM resonance contributions, where no search for dark photons is performed.
  The anti-$k_{\rm T}$-based isolation requirement is applied for $\ma > 1.1\gev$.
  }
  \label{fig:prompt_nexp}
\end{figure}

\vspace{1.5in}

\begin{figure}[h!]
  \centering
  \includegraphics[width=0.32\textwidth]{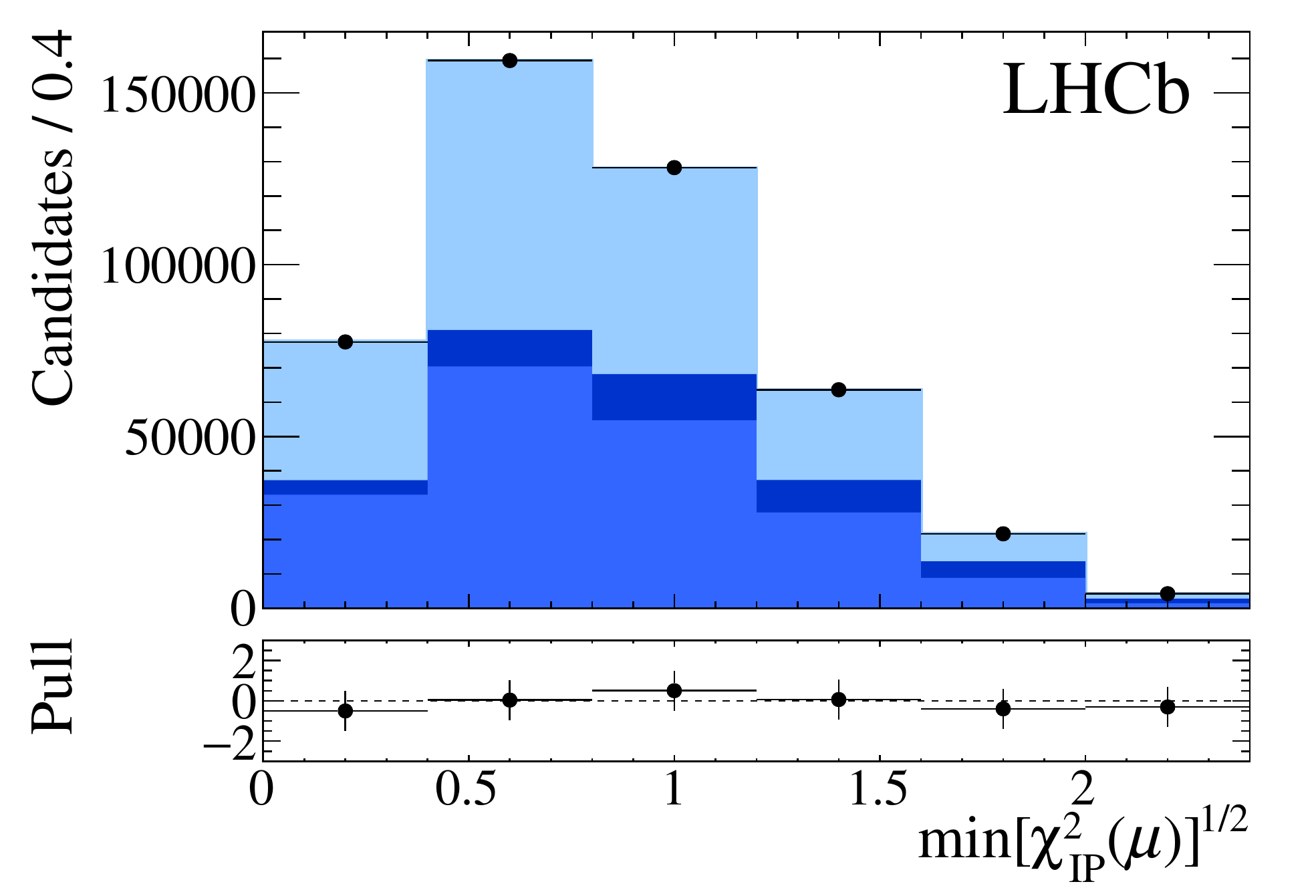}
  \includegraphics[width=0.32\textwidth]{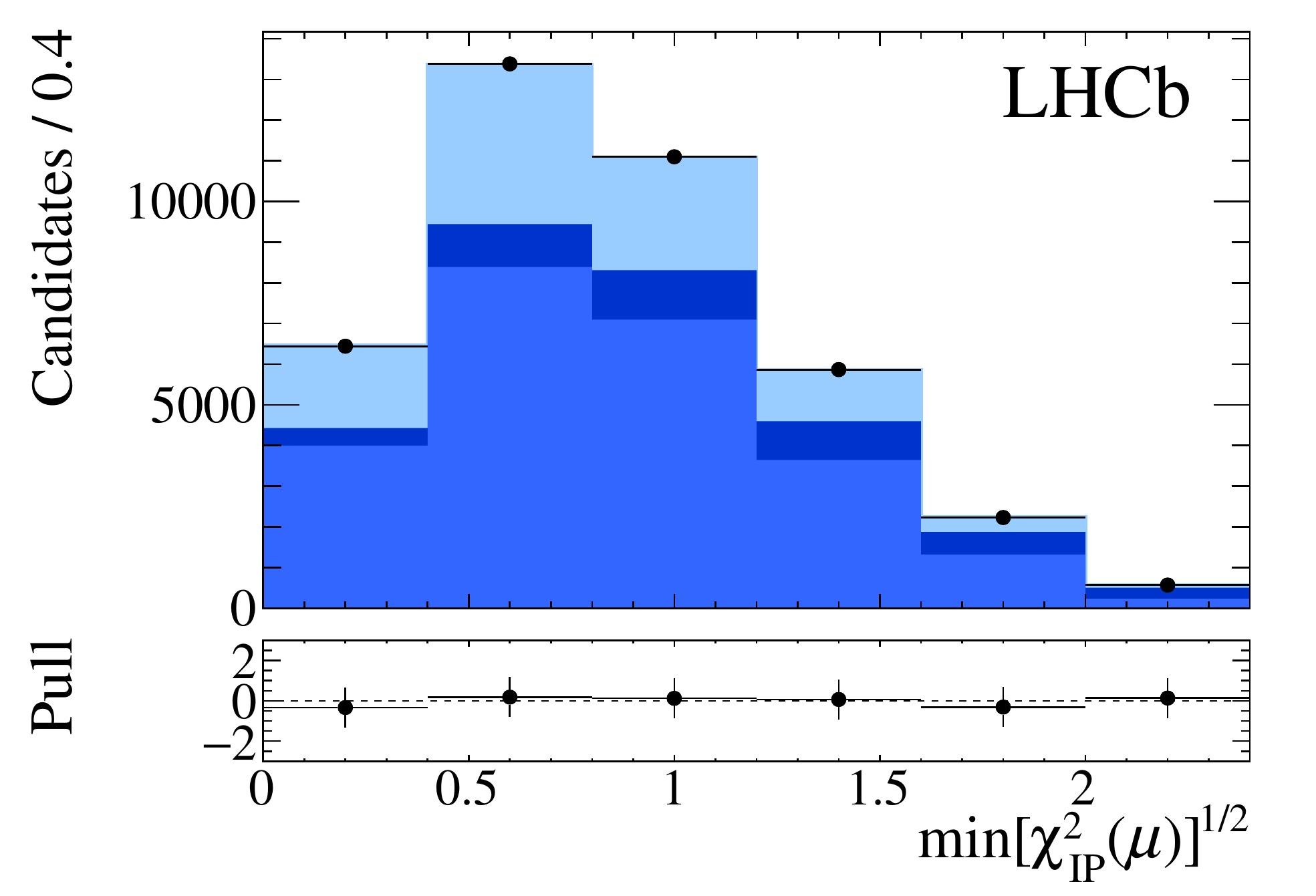}
  \includegraphics[width=0.32\textwidth]{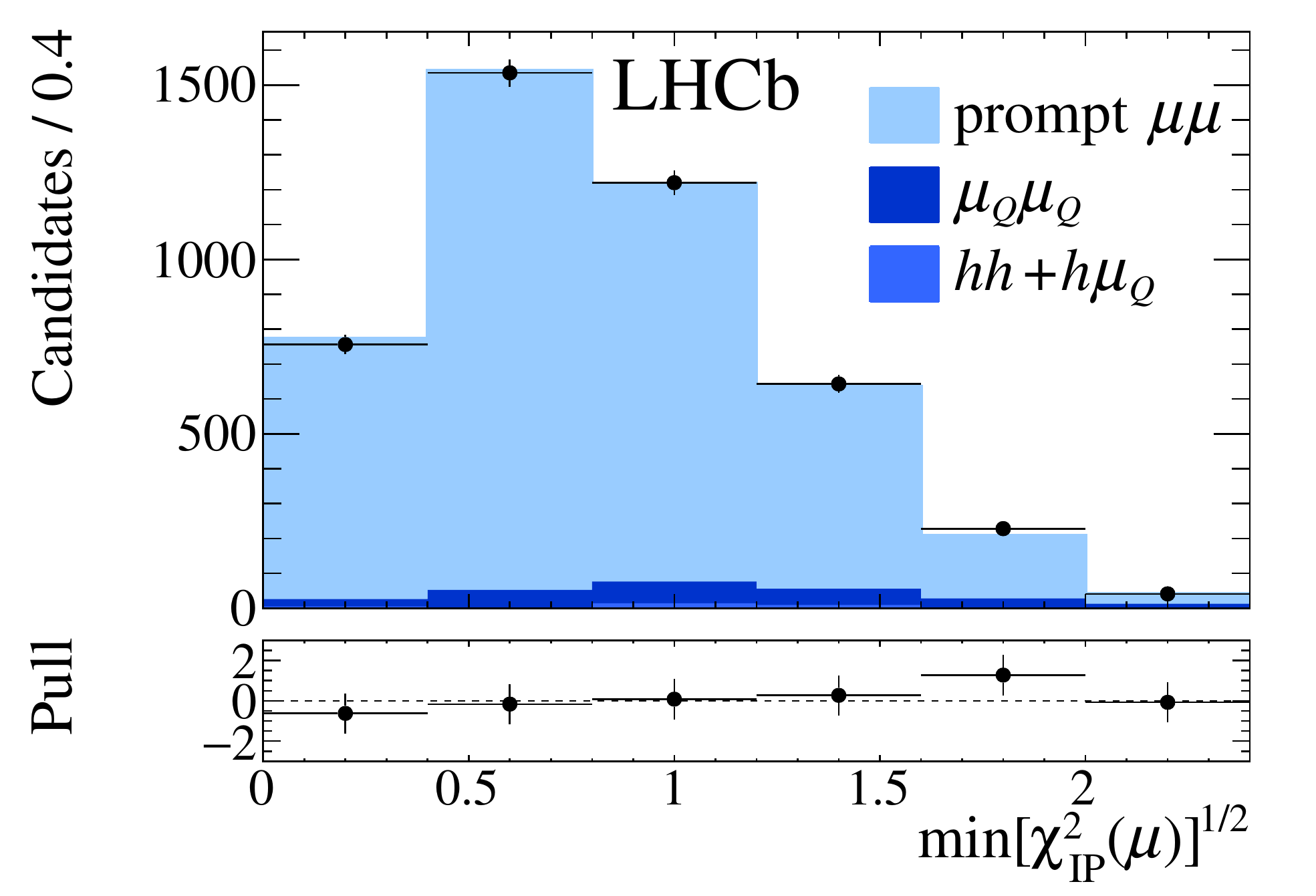}
  \caption{
  Example $\mxip^{1/2}$ distributions with fit results overlaid for prompt-like candidates near (left) $\ma=0.5$, (middle) 5, and (right) 50\gev. The square root of \mxip is used in the fits to increase the bin occupancies at large \mxip values.
  The background categories, as described in the Letter, are as follows: ($hh$) two prompt hadrons misidentified as muons;
($h\mu_Q$) a misidentified prompt hadron combined with a muon produced in the decay of a heavy-flavor quark, $Q$, that is misidentified as prompt;
and
($\mu_Q\mu_Q$) two muons produced in $Q$-hadron decays that are both  misidentified as prompt.
  }
  \label{fig:fits_prompt}
\end{figure}

\begin{figure}[h!]
  \centering
  \includegraphics[width=0.7\textwidth]{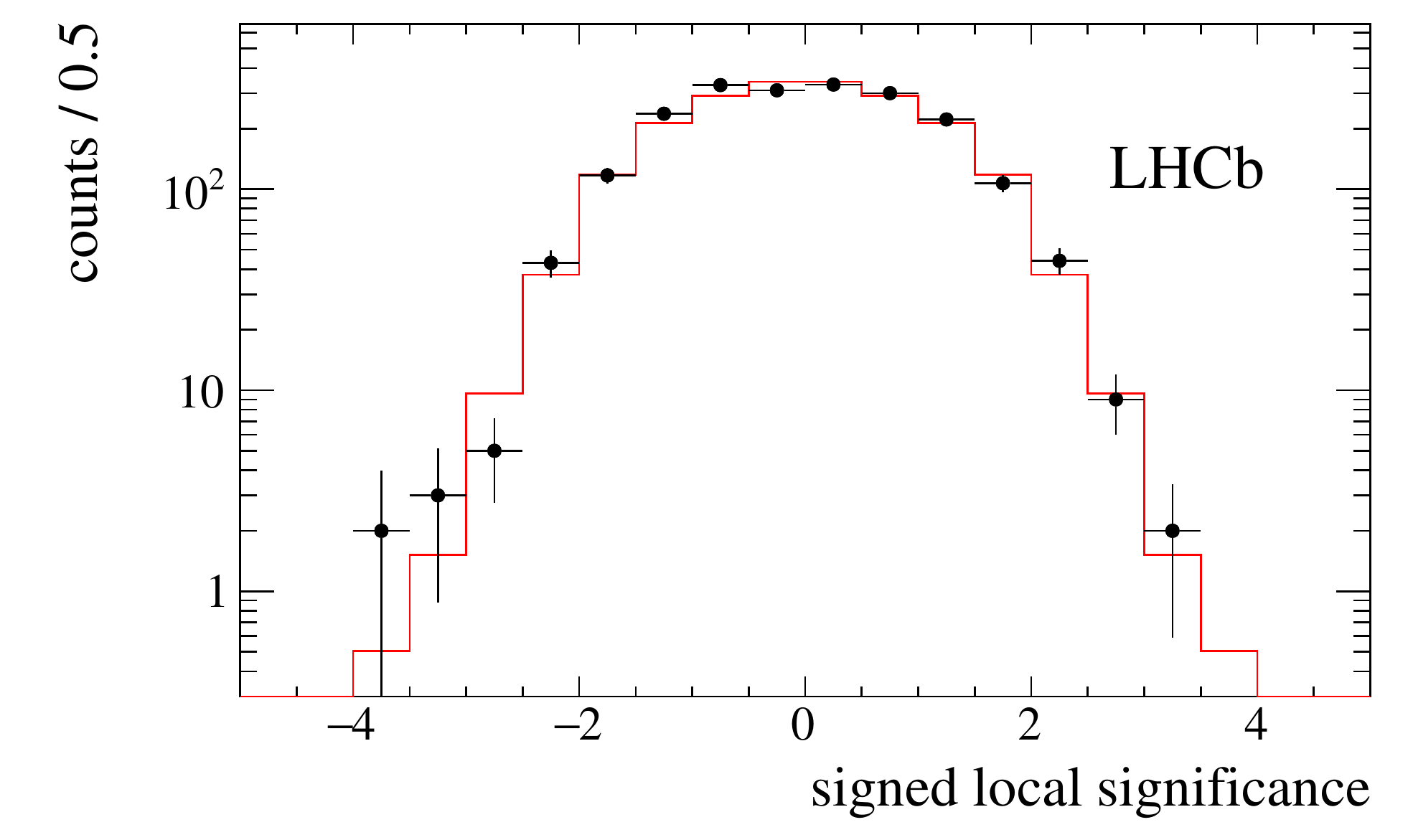}
\caption{ Signed local significances at each scan mass from (black points) all fits and (red) the expected distribution; if the best-fit signal-yield estimator is negative, the signed significance is negative and {\em vice versa}. The error bars account for the correlation between nearest-neighbor fit results, which often produces outliers in pairs due to the $\sa/2$ step size.
}
  \label{fig:prompt_pvals}
\end{figure}

\begin{figure}
  \centering
  \includegraphics[width=0.95\textwidth]{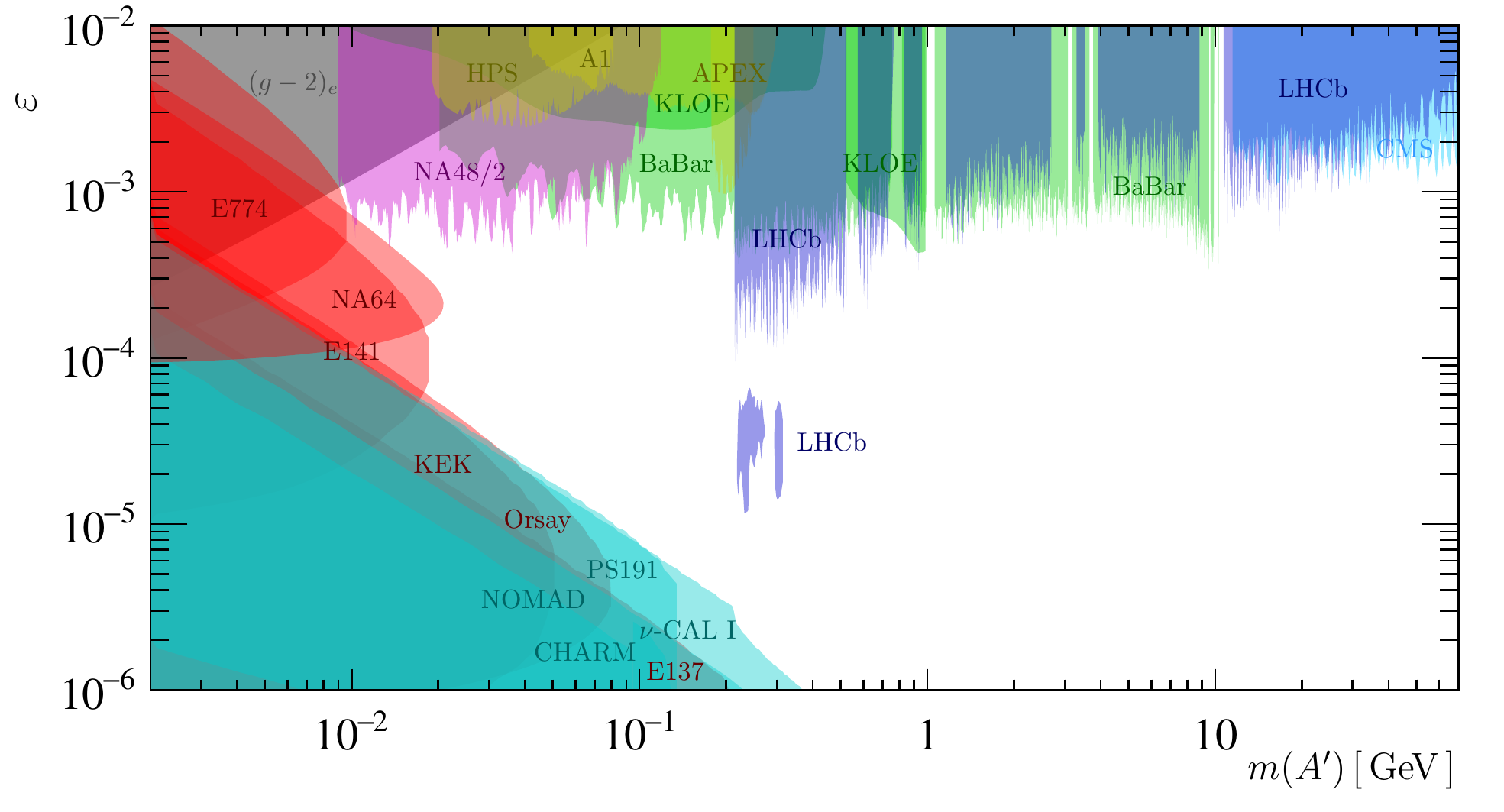}
  \caption{
  Comparison of the results presented in this Letter to existing constraints from previous experiments in the few-loop $\varepsilon$ region (see Ref.~\cite{Ilten:2018crw} for details about previous experiments).
  }
  \label{fig:all_lims}
\end{figure}

\begin{figure}
  \centering
  \includegraphics[width=0.95\textwidth]{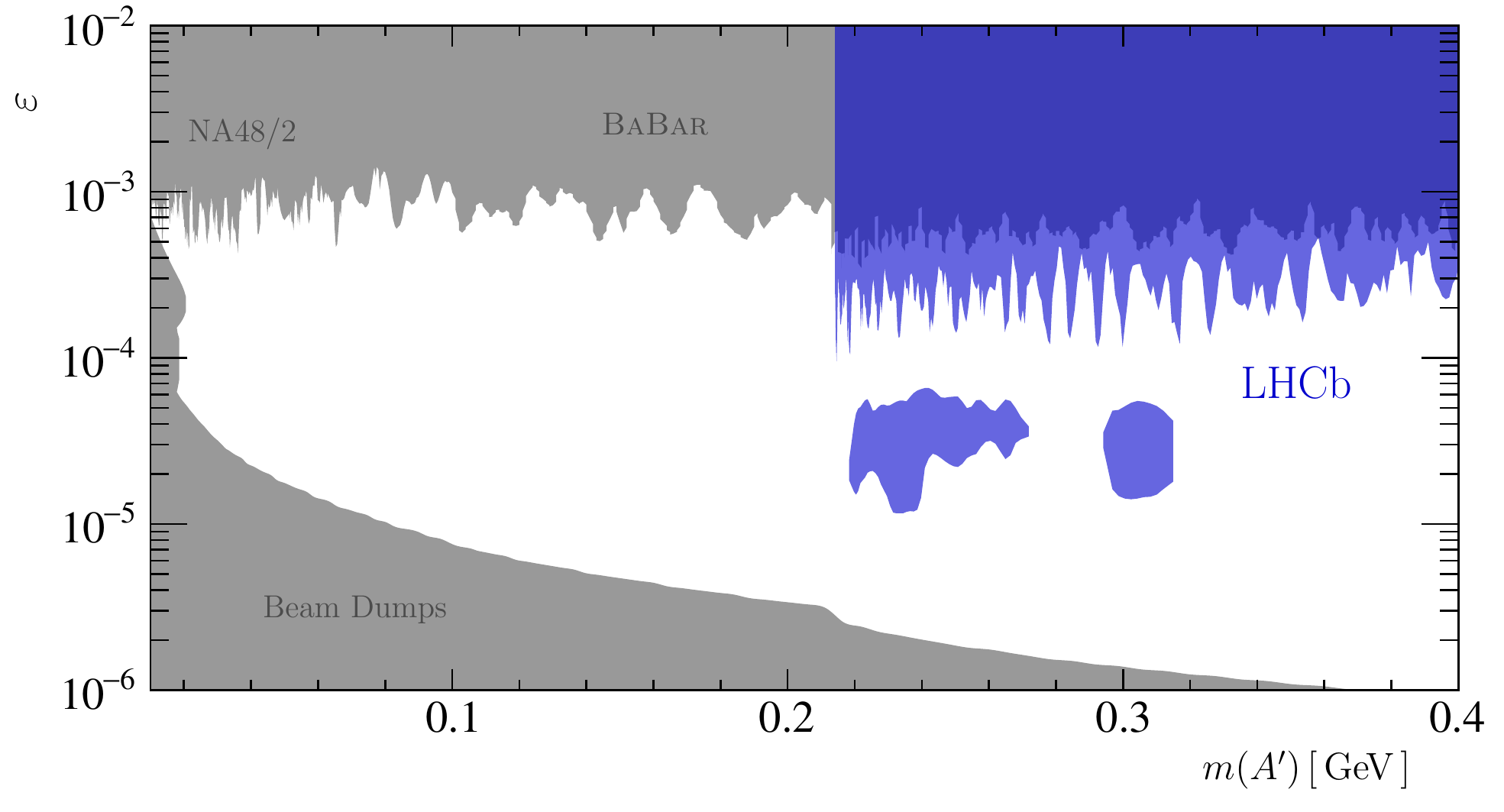}
  \caption{
  Comparison of the results presented in this Letter to existing constraints from previous experiments in the few-loop $\varepsilon$ region (see Ref.~\cite{Ilten:2018crw} for details about previous experiments), restricted to the mass region motivated by self-interacting dark matter~\cite{Tulin:2017ara}.
  }
  \label{fig:lims_lowmass}
\end{figure}

\clearpage

\setcounter{page}{1}
\pagenumbering{roman}

%\input{LHCb_Authorship_26-Aug-2019.tex}

% LHCb collaboration author list
% Data extracted on October 10th, 2019 at 3:33pm for reference date 26-Aug-2019
\centerline
{\large\bf LHCb collaboration}
\begin
{flushleft}
\small
R.~Aaij$^{31}$,
C.~Abell{\'a}n~Beteta$^{49}$,
T.~Ackernley$^{59}$,
B.~Adeva$^{45}$,
M.~Adinolfi$^{53}$,
H.~Afsharnia$^{9}$,
C.A.~Aidala$^{79}$,
S.~Aiola$^{25}$,
Z.~Ajaltouni$^{9}$,
S.~Akar$^{64}$
% ...

{\footnotesize \it

% ...
$ ^{9}$Universit{\'e} Clermont Auvergne, CNRS/IN2P3, LPC, Clermont-Ferrand, France\\
% ...
$ ^{25}$INFN Sezione di Milano, Milano, Italy\\
% ...
$ ^{31}$Nikhef National Institute for Subatomic Physics, Amsterdam, Netherlands\\
% ...
$ ^{45}$Instituto Galego de F{\'\i}sica de Altas Enerx{\'\i}as (IGFAE), Universidade de Santiago de Compostela, Santiago de Compostela, Spain\\
% ...
$ ^{49}$Physik-Institut, Universit{\"a}t Z{\"u}rich, Z{\"u}rich, Switzerland\\
% ...
$ ^{53}$H.H. Wills Physics Laboratory, University of Bristol, Bristol, United Kingdom\\
% ...
$ ^{59}$Oliver Lodge Laboratory, University of Liverpool, Liverpool, United Kingdom\\
% ...
$ ^{64}$University of Cincinnati, Cincinnati, OH, United States\\
% ...
$ ^{67}$Syracuse University, Syracuse, NY, United States\\
% ...
$ ^{79}$University of Michigan, Ann Arbor, United States, associated to $^{67}$\\
\bigskip
% ...
}
\end{flushleft}

\end{document}